\documentclass[12pt,fleqn]{article}

\usepackage{latexsym}

\usepackage[english]{babel}

\usepackage{amsmath,amssymb}
\usepackage{indentfirst}

\usepackage[dvips]{graphicx}

\begin{document}

\title{Possible imprints of cosmic strings in the shadows of galactic black holes}
\author{Vassil K. Tinchev\thanks{E-mail: tintschev@phys.uni-sofia.bg}, Stoytcho S. Yazadjiev\thanks{E-mail: yazad@phys.uni-sofia.bg}\\
{\footnotesize  Department of Theoretical Physics,
                Faculty of Physics, Sofia University,}\\
{\footnotesize  5 James Bourchier Boulevard, Sofia~1164, Bulgaria }\\}

\date{}

\maketitle

\begin{abstract}
We examine the shadow cast by a Kerr black hole pierced by a cosmic
string. The observable images depend not only on the black hole spin
parameter  and the angle of inclination, but also on the deficit
angle yielded by the cosmic string. The dependence of the observable
characteristics of the shadow on the deficit angle is explored. The
imprints in the black hole shadow left by the presence of a cosmic
string can serve as a method for observational detection of such
strings.
\end{abstract}

It's well known that the shadow (or the apparent shape) of a compact
relativistic object encodes  information about the nature of this
object \cite{FalckeEtAll}. That is why the apparent shapes of
various black holes and other compact objects, such as wormholes and
naked singularities, have been intensively studied in the last
years. The shadows of the black holes and naked singularities  from
the Kerr-Newmann family of solutions of the Einstein-Maxwell
equations have been thoroughly investigated in
\cite{Bardeen}-\cite{bibl2}. The shadow of a black hole with a
NUT-charge has been obtained in \cite{AbdujabbarovEtAll}. The black
hole shadows in Einstein-Maxwell-dilaton gravity,  Chern-Simons
modified gravity and braneworld gravity  have been examined in
\cite{AmarillaEiroa}, \cite{AmarillaEiroaGiribet}, \cite{AmarillaEiroa1}.
The apparent shape of the Sen black hole has been studied in
\cite{HiokiMiyamoto}. The wormhole shadows have been recently
investigated in \cite{BambiFreese}, \cite{NedkovaTinchevYazadjiev}.

With the advance of  technology the experimental observation of the
shadows of  compact objects is now possible. Experiments that allow
such observations include the Event Horizon Telescope
\cite{EHTelescope}, which is a system of earth-based telescopes
measuring in the (sub)millimeter wavelength, the space-based radio
telescopes RadioAstron and Millimetron \cite{Radioastron},
\cite{JohannsenEtAll}, and the space-based X-ray interferometer
MAXIM \cite{MAXIM}. In the next few years these missions are
expected to reach resolution high enough to observe the shadow of
the supermassive compact object at the center of our galaxy or those
located at nearby galaxies \cite{JohannsenEtAll}. The results of
these experiments should be compared with the theoretical models. In
this way the observations will reject some of the models or will
make it possible to distinguish between different types of compact
objects \cite{LiBambi}. Even more, the mentioned observations can be
used for detection of theoretically predicted objects and effects
not observed so far. Such a theoretical prediction is the possible
existence of cosmic strings. These objects are expected to have
formed during phase transitions in the early universe through
spontaneous symmetry breakings \cite{bibl0}. It has also been shown
that cosmic strings generally form at the end of inflation within
the framework of various supersymmetric grand unification theories
\cite{Jeannerot2003}.

A cosmic string makes the space-time around it  a conical space-time
with a deficit angle $\delta= 8\pi G\mu/c^4$ where $\mu$ is the
string tension and $G$ and $c$ are the gravitational constant and
speed of light, respectively. The deficit angle manifests itself
physically by giving rise to interesting phenomena and effects
\cite{bibl0}, \cite{GaltsovMasar}-\cite{bibl3} which can be used for
detection of  cosmic strings. For example, the gravitational lensing
phenomena serve as direct evidence for cosmic strings although none
have been detected yet. However, it is unlikely that the pure
gravitational lensing of a single string could be measured in the
violent astrophysical conditions. It is much more natural to search
for signs of cosmic strings in the gravitational lensing by the
whole system consisting of the cosmic string and the matter
surrounding it. In a galactic context we can consider a model
configuration consisting of a central galactic black hole pierced by
a cosmic string. The gravitational lensing of such configurations
has been recently investigated in \cite{bibl3} and it has been shown
that the presence of a cosmic string leaves observable imprints.
Then, in view of these results, it is very natural to expect that
the presence of a cosmic string would also leave imprints in the
shadow of the black hole pierced by it.

With this motivation in mind, we continue the theoretical analysis
of the black hole shadows. The aim of the current paper is to
investigate the apparent shape of a rotating black hole pierced by a
cosmic string, and compare the results with the case of the Kerr
black hole. Then we consider the deviations from the Kerr case as a
possible test for the existence of cosmic strings on a galactic
level.

The metric of the spacetime describing a rotating Kerr black hole
pierced along  the axis of symmetry by a cosmic string\footnote{For
numerical solutions describing rotating black holes with cosmic
string hair we refer the reader to \cite{Gregory2013}. } is given by
\cite{GaltsovMasar}

\begin{equation}\label{metric}
  ds^2=-\left(1-\frac{2Mr}{\rho^{2}}\right)dt^2+\rho^{2}\left(\frac{dr^2}{\Delta}\ +
  d\theta^{2}\right)+\zeta\ \frac{\sin^2\theta}{\rho^{2}}\left(\zeta\ \Sigma^{2}d\varphi-4aMr\
  dt\right)d\varphi,
\end{equation}
where  $r$, $\theta$ and $\varphi$ are the Boyer-Lindquist
coordinates and the metric functions $\Delta$, $\rho$ and $\Sigma$
are defined  as usual

\begin{equation}\label{metricFunc}
  \Delta\equiv r^2-2Mr+a^2,\ \rho^2\equiv r^2+a^2\cos^2\theta,\ \Sigma^2\equiv
  \left(a^2+r^2\right)^{2}-a^2\Delta\sin^2\theta,
\end{equation}
where $M$, $a$ and $\zeta$ are parameters. The parameter $\zeta$
($0<\zeta\leq 1$)  describes the influence of the string on the
metric and it is related to the deficit angle of the string by
$\delta=2\pi(1-\zeta)$. In the particular case when $\zeta=1$, the
metric reduces to the well-known  metric of Kerr. The parameter $a$
is the angular momentum per unit mass. $M$, however,  does not
coincide with the physical mass of the black hole.  The physical
mass $M_{phys}$ and the physical angular momentum $J_{phys}$ of the
black hole pierced by a cosmic string are given by
\cite{GaltsovMasar}

\begin{eqnarray}\label{physpar}
M_{phys}=\zeta M, \; \; \; J_{phys}= \zeta J.
\end{eqnarray}

The motion of  test particles in spacetime is determined  by the
geodesic equations or equivalently by the Hamilton-Jacobi equation

\begin{equation}\label{HamiltonJacobiEq}
-2\frac{\partial S}{\partial \lambda}= g^{\alpha\beta}\frac{\partial
S}{\partial x^{\alpha}}\frac{\partial S}{\partial x^{\beta}},
\end{equation}
where $S$ is the particle action, $\lambda$ is the affine parameter
along the geodesics of the metric $g_{\alpha\beta}$. Since our
spacetime described by the metric (\ref{metric}) is stationary and
axisymmetric we have two conserved quantities  - the energy of the
particle $E$ and its angular momentum $L_{z}$ about the axis of
symmetry. As in the case of  pure Kerr spacetime,  we have another
conserved quantity, namely the Carter constant $K$,  leading to the
separability of the Hamilton-Jacobi equation, which then  has a
solution of the form

\begin{equation}\label{solHamiltonJacobiEq}
  S=\frac{1}{2}m^{2}\lambda-Et+L_{z}\varphi+S_{r}(r)+S_{\theta}(\theta),
\end{equation}
where $m$ is the mass of the test particle.  Using
(\ref{solHamiltonJacobiEq}) the Hamilton-Jacobi equation reduces to
the following equations for $S_{r}(r)$ and $S_{\theta}(\theta)$:

\begin{equation}\label{soleqrtheta1}
    \begin{array}{ll}
      \left(S'_{r}\right)^{2}=\frac{\left(a^{2}+r^{2}\right)^{2}}{\Delta^{2}}\ E^{2}
      +\frac{a^{2}}{\zeta^{2}\Delta^{2}}\ L_{z}^{2}-\frac{4aMr}{\zeta\Delta^{2}}\ EL_{z}-\frac{m^{2}r^{2}+K}{\Delta}\equiv R(r),\\
      \\
      \left(S'_{\theta}\right)^{2}=K-m^{2}a^{2}\cos^{2}\theta-E^{2}a^{2}\sin^{2}\theta-
      \frac{1}{\zeta^{2}\sin^{2}\theta}\ L_{z}^{2}\equiv
      \Theta(\theta).
    \end{array}
\end{equation}

Then \eqref{solHamiltonJacobiEq} takes the form
\begin{equation}\label{solHamiltonJacobiEqConcrete}
    S=\frac{1}{2}m^{2}\lambda-Et+L_{z}\varphi+\int \sqrt{R(r)}\ dr+\int \sqrt{\Theta(\theta)}\ d\theta.
\end{equation}

Hence by using  the standard procedure   we find the null geodesics
(i.e. $m=0$) in the spacetime of a rotating black hole pierced by a
cosmic string, namely
\begin{equation}\label{eqGeodesics1AIsotropic}
    \begin{array}{ll}
        \rho^{2}\ \frac{dt}{d\lambda}=\frac{1}{\Delta}\left[\left(a^{2}+r^{2}\right)^{2}-
    \frac{2aMr}{\zeta}\ \xi\right]-a^{2}\sin^{2}\theta,\\
        \\
        \rho^{2}\ \frac{dr}{d\lambda}=\sqrt{\left(a^{2}+r^{2}\right)^{2}+
    \frac{a^{2}}{\zeta^{2}}\ \xi^{2}-\frac{4aMr}{\zeta}\ \xi-\Delta\eta}\equiv\sqrt{R},\\
        \\
        \rho^{2}\ \frac{d\theta}{d\lambda}=\sqrt{\eta-a^{2}\sin^{2}\theta-
      \frac{1}{\zeta^{2}\sin^{2}\theta}\ \xi^{2}}\equiv\sqrt{\Theta},\\
        \\
        \zeta\rho^{2}\ \frac{d\varphi}{d\lambda}=\frac{\xi}{\sin^{2}\theta}-
    \frac{a}{\Delta}\left(\frac{a}{\zeta}\ \xi-2Mr\right),
    \end{array}
\end{equation}
with   $\xi\equiv L_{z}/E$ and $\eta\equiv K/E^{2}$ being the impact
parameters. We have also redefined the affine parameter  $E\lambda
\to \lambda$.

The photon orbits are in general of two types - orbits falling into
the black hole and others scattered away from the black hole to
infinity. An observer far from the black hole will be able to see
only the photons scattered away from the black hole, while those
captured by the black hole will form a dark region. This dark region
observed on the luminous background  is  the shadow of the black
hole.

The boundary of the black hole shadow is the critical orbit that
separates the escape and plunge orbits. In order to find the shadow
boundary we reformulate the problem as one-dimensional problem for a
particle  in an effective potential by rewriting the radial geodesic
equation in the form

\begin{equation*}
  \left(\rho^{2}\
  \frac{dr}{d\lambda}\right)^{2}+U_{eff}(r)=0,
\end{equation*}
where $U_{eff}(r)=-R(r)$. In this formulation, it is clear that the
critical orbit between escape and plunge, which is obviously
unstable and circular,  corresponds to the highest maximum of the
effective potential. The conditions therefore for the critical
spherical orbit determining the boundary of the black hole shadow
are

\begin{eqnarray}\label{CONU}
U_{eff}=0, \;\; \frac{U_{eff}}{dr}=0, \;\;
\frac{d^2U_{eff}}{dr^2}\le 0,
\end{eqnarray}
or, equivalently, $R=0$, $\frac{dR}{dr}=0$ and $\frac{d^2R}{dr^2}\ge
0$. Since the effective potential $U_{eff}$ (or equivalently $R$)
depends on $r$ as well as  $\xi$ and $\eta$, the conditions
(\ref{CONU}) give in fact a parametric relation between the impact
parameters that should be satisfied on the shadow boundary. Of
course, in addition to the  conditions above, the impact parameters
should be such that $\Theta(\theta)\ge 0$.

Taking into account the explicit form of the function $R$ in our
case, the solution to the conditions $R=0$ and $\frac{dR}{dr}=0$,
that also satisfies $\Theta(\theta)\ge 0$, is given by

\begin{equation}\label{solxieta2}
    \begin{array}{ll}
    \xi=-\frac{\zeta}{a(r-M)}\left[r^2(r-3M) + a^2(r+M)\right],\\
    \\
    \eta=\frac{2}{(r-M)^{2}}\left[r^{2}(r^2-3M^2)+a^2(r^2+M^2)\right].
    \end{array}
\end{equation}

The condition  $\frac{d^2R}{dr^2}\ge 0$  leads to the following
explicit  inequality
\begin{equation}\label{d2Rrsolxieta2}
    3r^{2}+a^{2}-\frac{1}{(r-M)^{2}}\left[r^{2}(r^2-3M^2)+a^2(r^2+M^2)\right]\ge
    0.
\end{equation}

Equations (\ref{solxieta2}) and (\ref{d2Rrsolxieta2}) define the
boundary of the shadow in  parametric form. It is clear from the
derivation that the boundary of the shadow is determined only by the
spacetime  metric and does not depend on the details of the emission
mechanisms.

In  real observations, however, what is seen, is in fact the
projection of the shadow on the observer's sky defined as the plane
passing through the black hole and normal to the line of sight.
Taking this into account, it is more natural to present the shadow
boundary in the so-called celestial coordinates $\alpha$ and
$\beta$. The celestial coordinates are defined by \cite{bibl1}

\begin{equation}\label{limalpha}
    \alpha= \lim_{r\to\ \infty}\left(-r^{2}\sin\theta_{0}\frac{d\varphi}{dr}\right),
\end{equation}
\begin{equation}\label{limbeta}
  \beta=\lim_{r\to\ \infty}\left(r^{2}\frac{d\theta}{dr}\right),
\end{equation}
where the limit is taken along the null geodesics and  $\theta_{0}$
is the inclination angle between the axis of rotation of the black
hole and the line of sight of the observer. From the definition of
the celestial coordinates and using the null geodesics equations
\eqref{eqGeodesics1AIsotropic}, we get
\begin{equation}\label{alphabetaA}
    \begin{array}{ll}
  \alpha=-\frac{\xi}{\zeta\sin\theta_{0}}\ ,\\
  \\
  \beta=\sqrt{\eta-a^{2}\sin^{2}\theta_{0}-\frac{\xi^{2}}{\zeta^{2}\sin^{2}\theta_{0}}}\ .
    \end{array}
\end{equation}
After substituting \eqref{solxieta2} in \eqref{alphabetaA}, we find
\begin{equation}\label{solalphabetaA2}
    \begin{array}{ll}
    \alpha=\frac{1}{a(r-M)\sin\theta_{0}}\left[r^2(r-3M) + a^2(r+M)\right],\\
    \\
    \beta=\sqrt{\frac{2\left[r^{2}(r^2-3M^2)+a^2(r^2+M^2)\right]}{(r-M)^{2}}-a^{2}\sin^{2}\theta_{0}-
      \frac{\left[r^2(r-3M) + a^2(r+M)\right]^{2}}{a^{2}(r-M)^{2}\sin^{2}\theta_{0}}}.
    \end{array}
\end{equation}
These two equations give in parameteric form the boundary of the
shadow in celestial coordinates. Formally (\ref{solalphabetaA2})
coincide with the celestial coordinates for the shadow of the pure
Kerr black hole. However, although the string parameter $\zeta$ does
not enter  explicitly the equations for the shadow boundary in
celestial coordinates, it indeed  influences the black hole shadow
through the parameter $M$, which is related to the physical mass of
the black hole via eq. (\ref{physpar}), i.e. $M_{phys}=\zeta M$.

\begin{figure}[h]
        \setlength{\tabcolsep}{ 0 pt }{\scriptsize\tt
        \begin{tabular}{ cccc }
            \includegraphics[width=4.1cm]{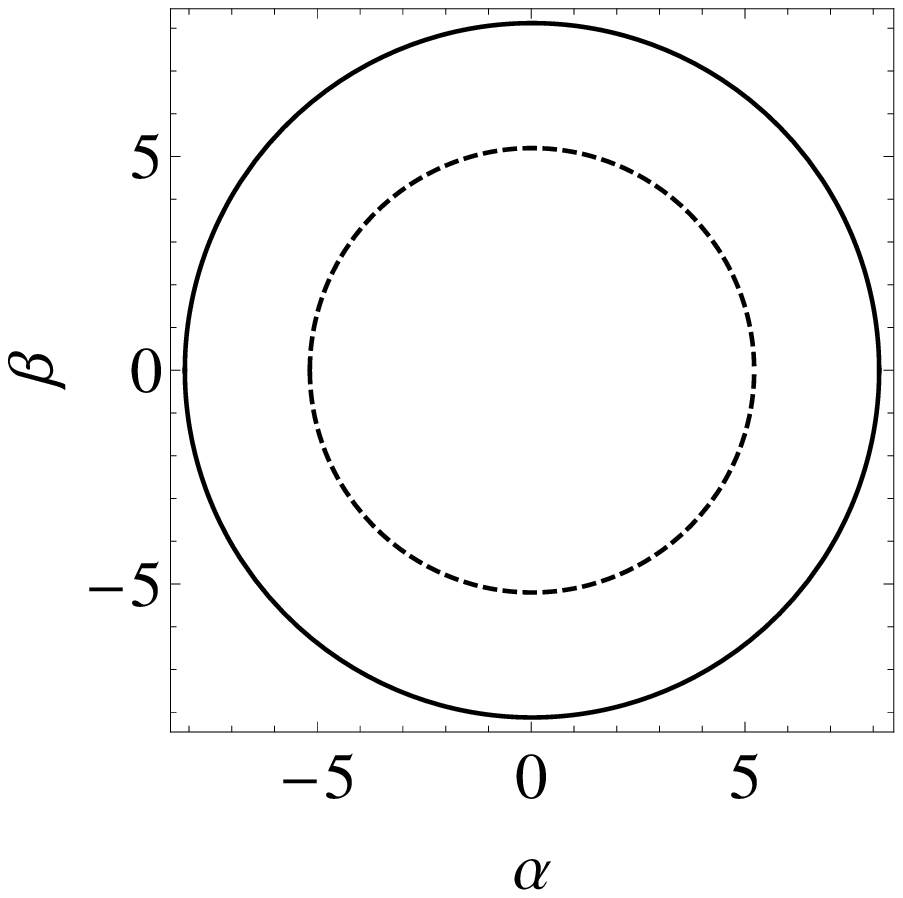} &
            \includegraphics[width=4.1cm]{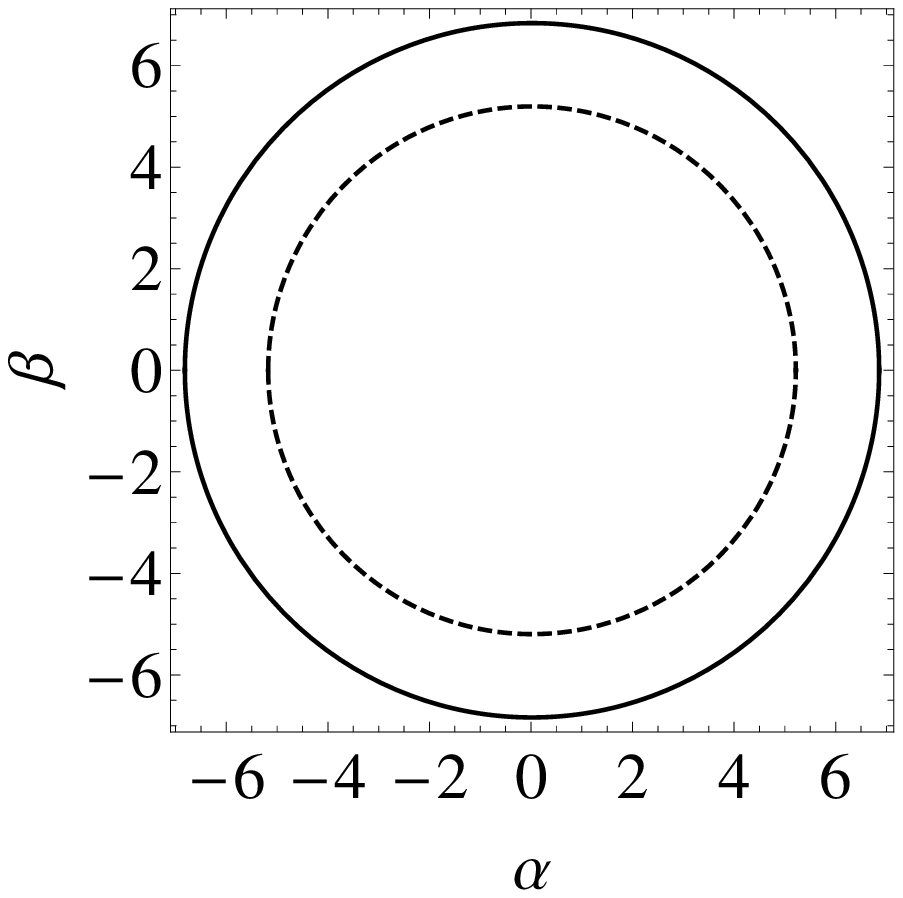} &
            \includegraphics[width=4.1cm]{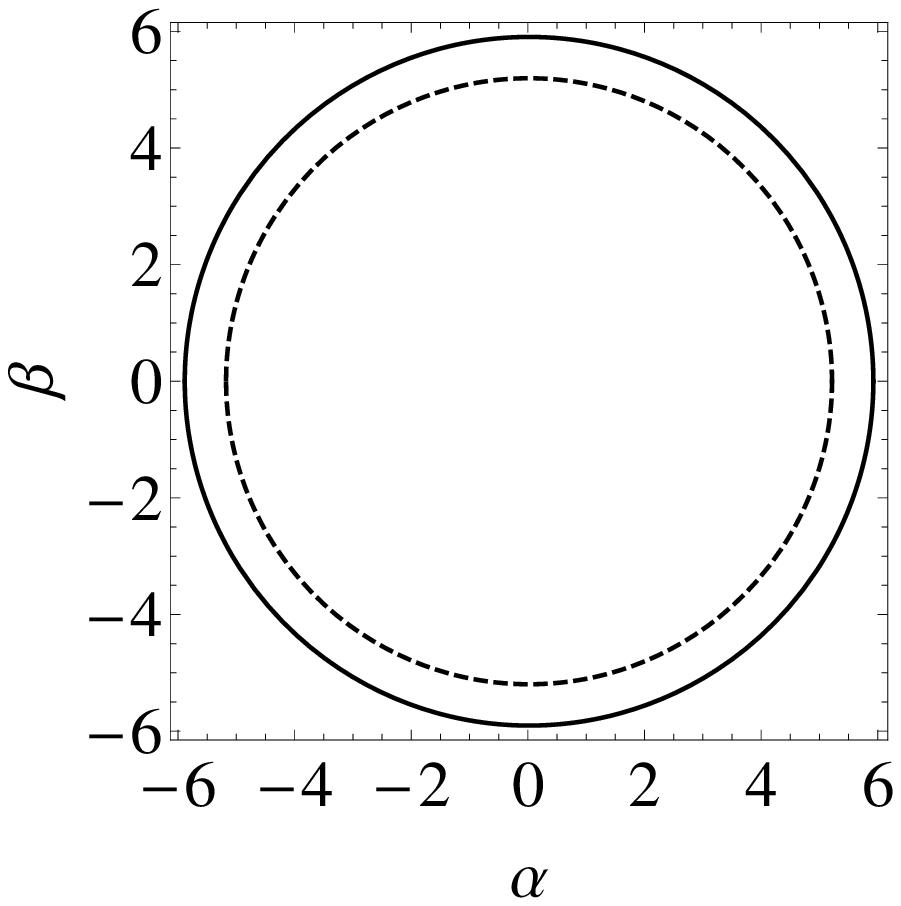} &
            \includegraphics[width=4.1cm]{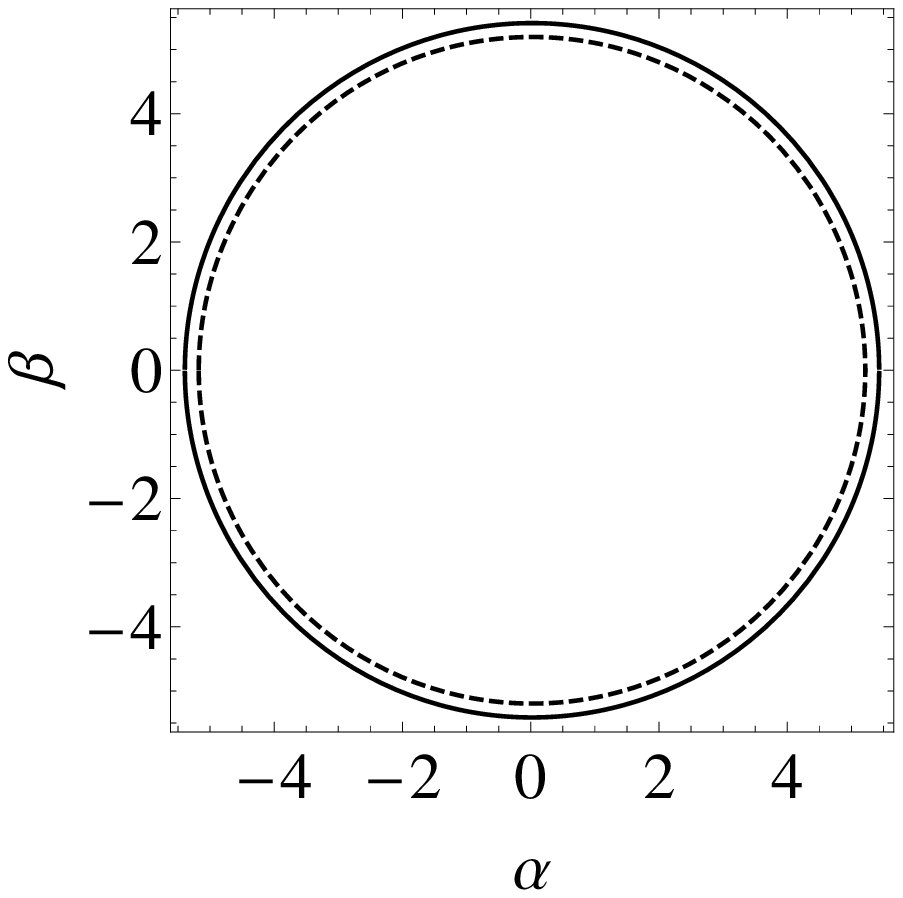} \\
            $a/M_{phys}=0.01$, $\zeta=0.64$;\  &
            $a/M_{phys}=0.01$, $\zeta=0.76$;\  &
            $a/M_{phys}=0.01$, $\zeta=0.88$;\  &
            $a/M_{phys}=0.01$, $\zeta=0.96$ \\
            \includegraphics[width=4.1cm]{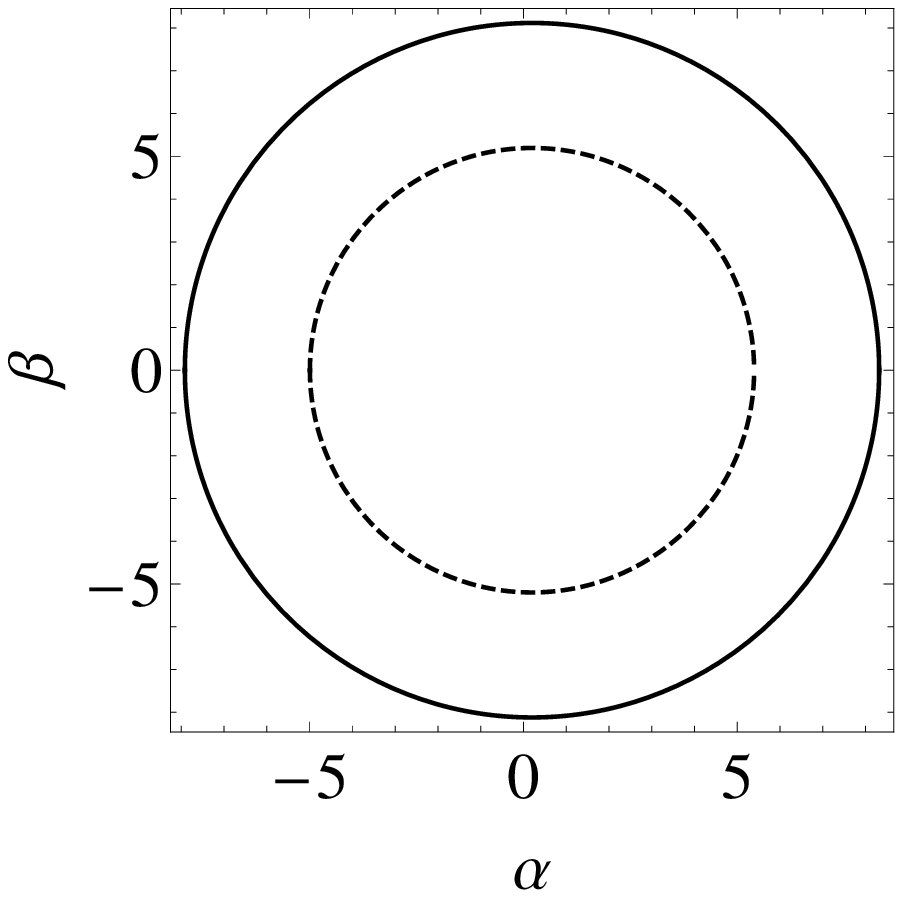} &
            \includegraphics[width=4.1cm]{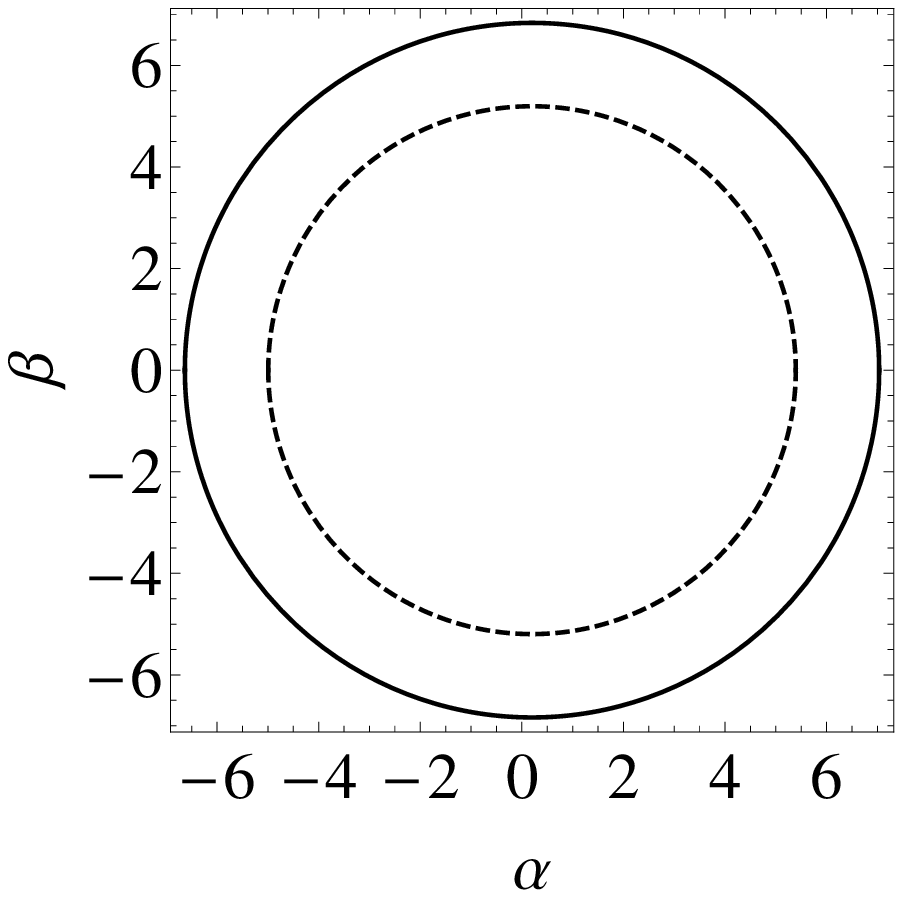} &
            \includegraphics[width=4.1cm]{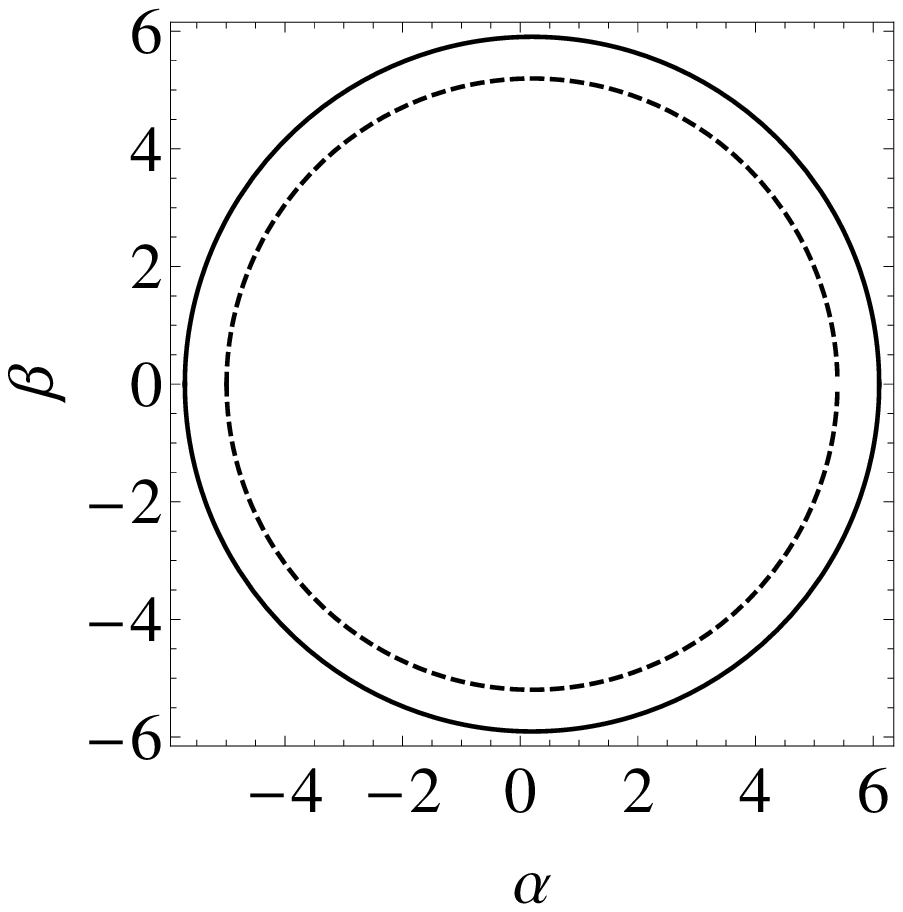} &
            \includegraphics[width=4.1cm]{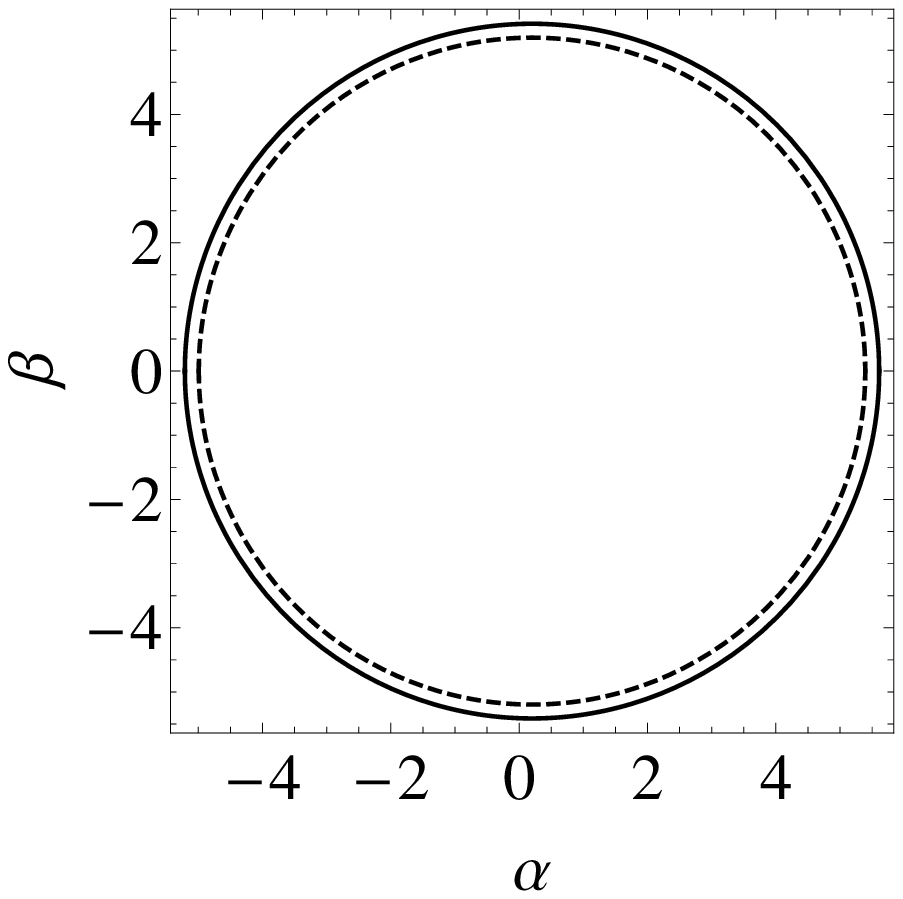} \\
            $a/M_{phys}=0.1$, $\zeta=0.64$;\  &
            $a/M_{phys}=0.1$, $\zeta=0.76$;\  &
            $a/M_{phys}=0.1$, $\zeta=0.88$;\  &
            $a/M_{phys}=0.1$, $\zeta=0.96$ \\
            \includegraphics[width=4.1cm]{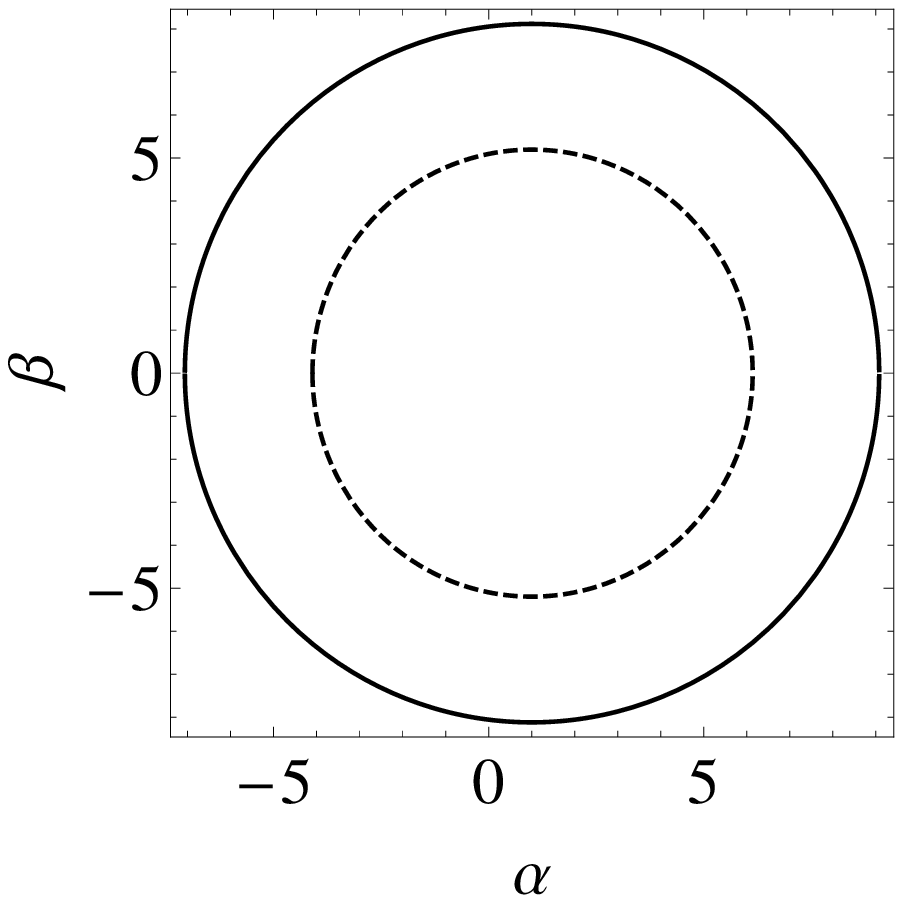} &
            \includegraphics[width=4.1cm]{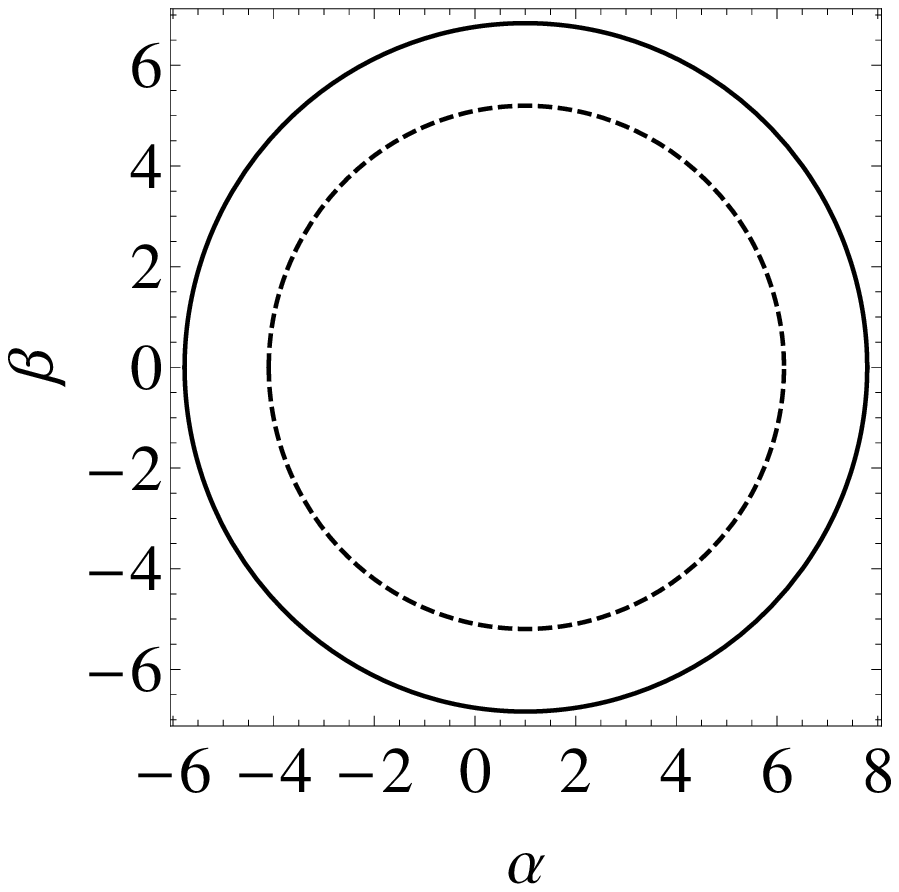} &
            \includegraphics[width=4.1cm]{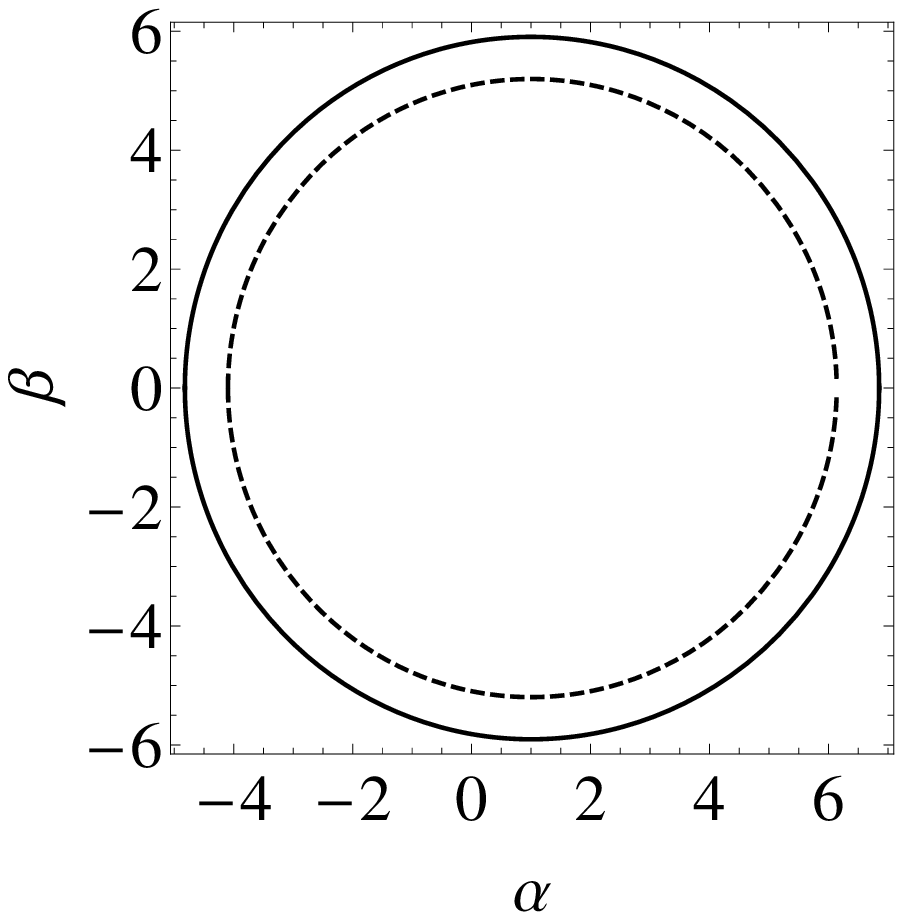} &
            \includegraphics[width=4.1cm]{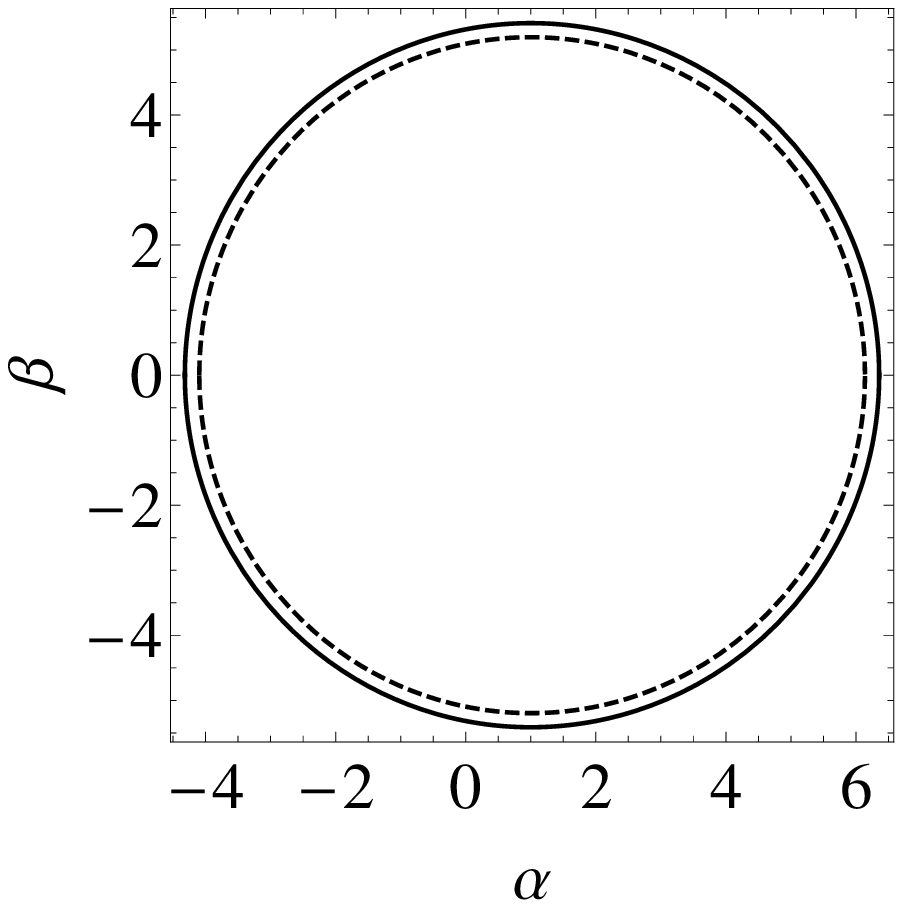} \\
            $a/M_{phys}=0.5$, $\zeta=0.64$;\  &
            $a/M_{phys}=0.5$, $\zeta=0.76$;\  &
            $a/M_{phys}=0.5$, $\zeta=0.88$;\  &
            $a/M_{phys}=0.5$, $\zeta=0.96$ \\
            \includegraphics[width=4.1cm]{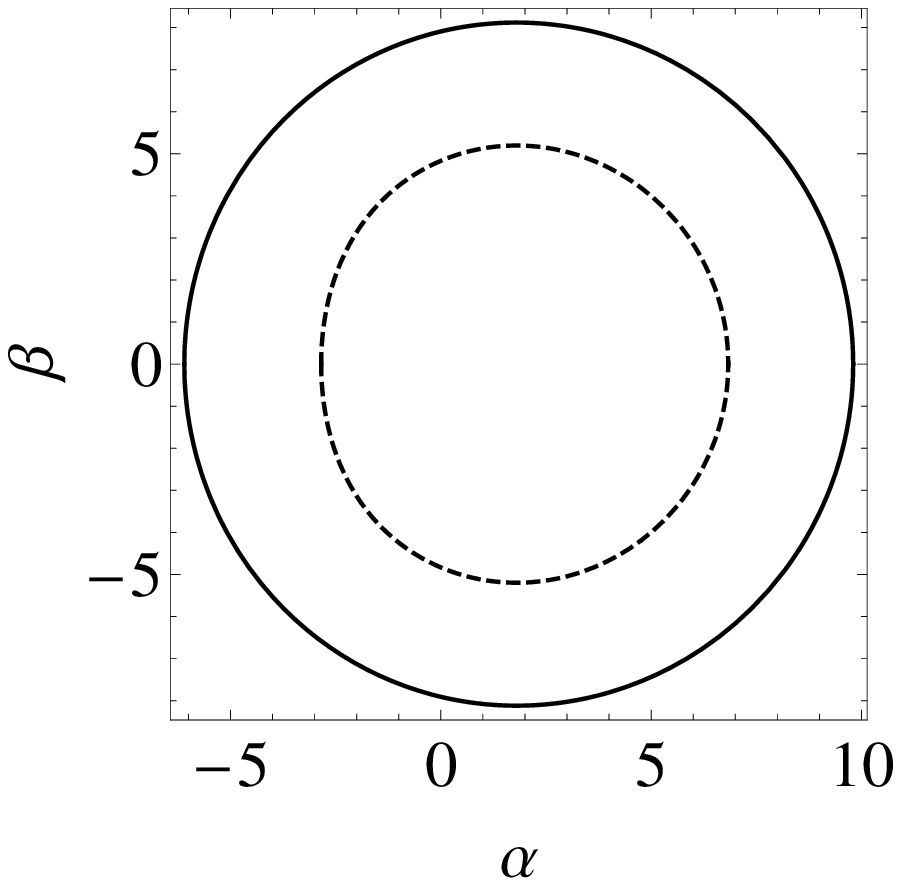} &
            \includegraphics[width=4.1cm]{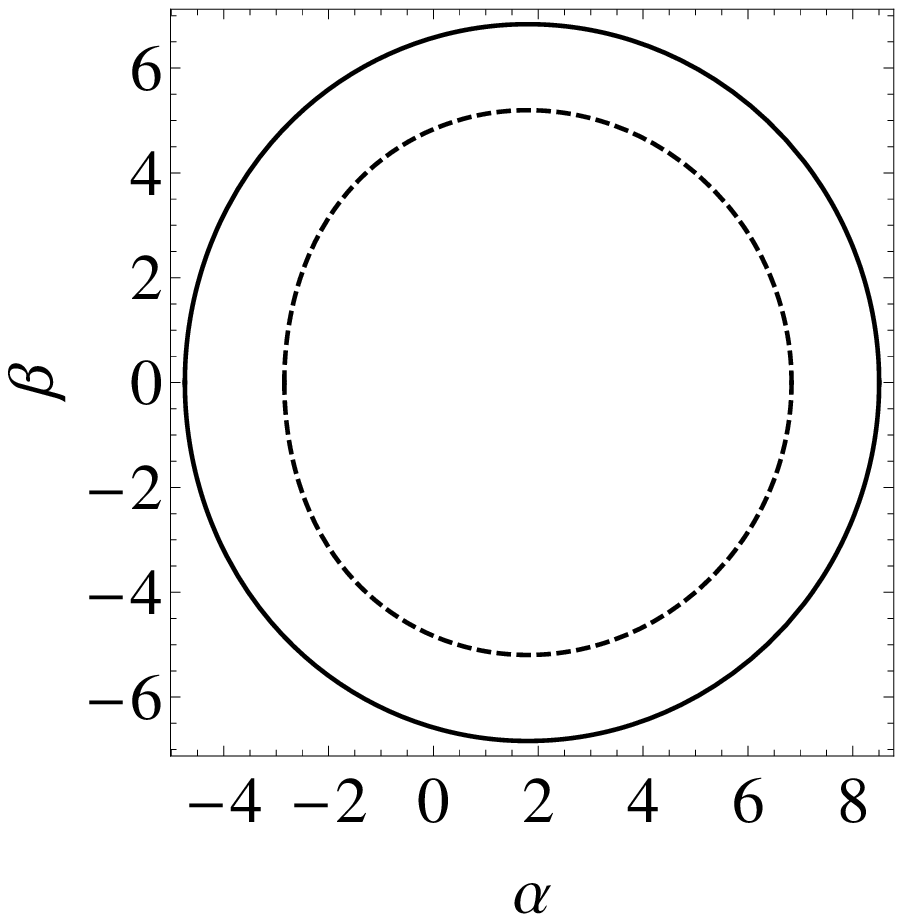} &
            \includegraphics[width=4.1cm]{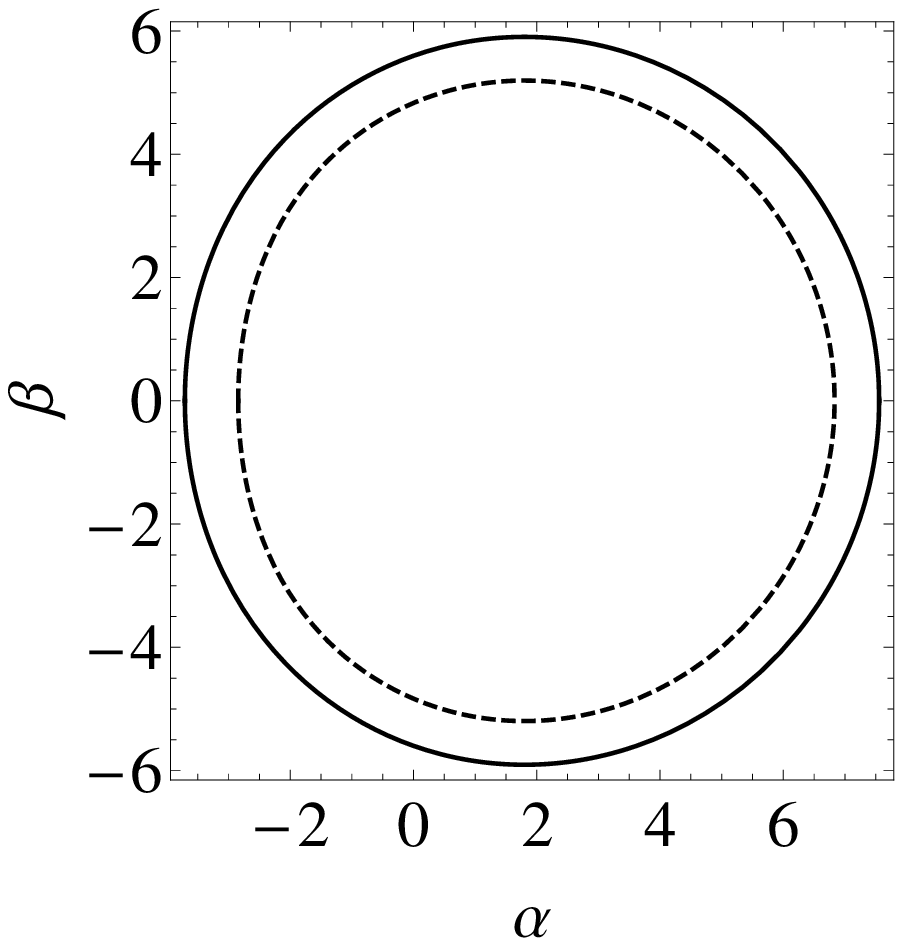} &
            \includegraphics[width=4.1cm]{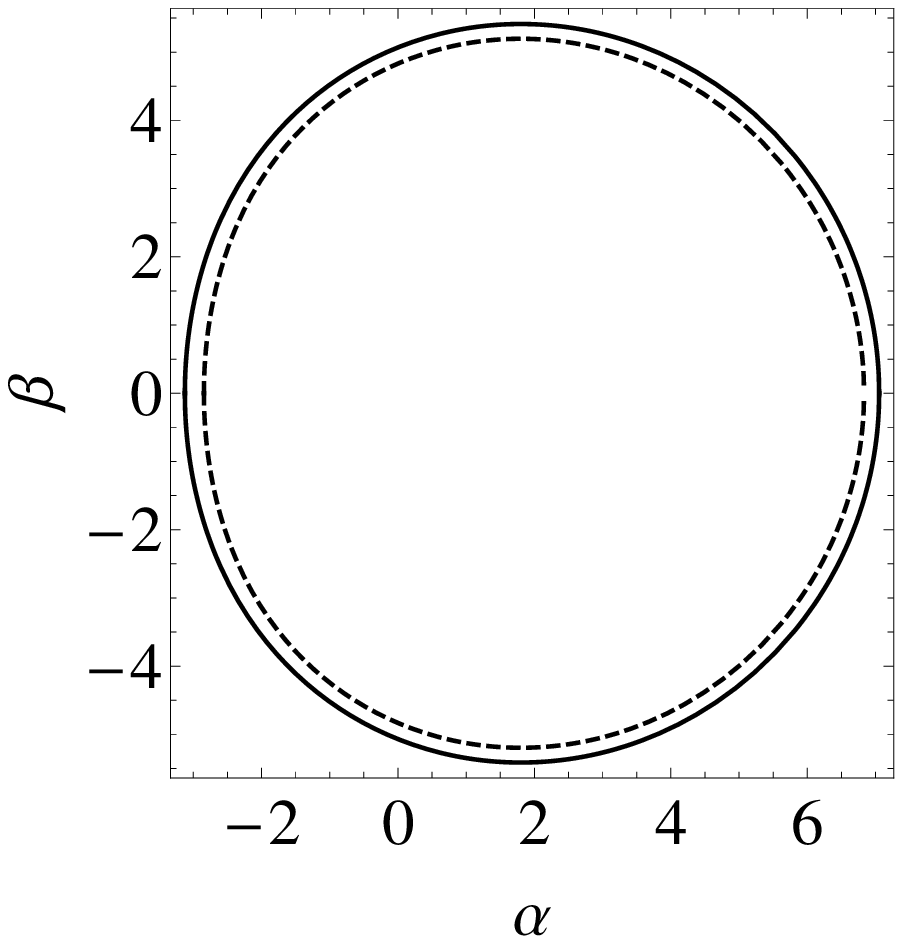} \\
            $a/M_{phys}=0.9$, $\zeta=0.64$;\  &
            $a/M_{phys}=0.9$, $\zeta=0.76$;\  &
            $a/M_{phys}=0.9$, $\zeta=0.88$;\  &
            $a/M_{phys}=0.9$, $\zeta=0.96$ \\
        \end{tabular}}
\caption{\footnotesize{The shadow of the Kerr black hole pierced by
a cosmic string (solid line) and the Kerr black hole (dashed line)
with inclination angle $\theta_{0}=\pi/2\ rad$ for different
rotation and string parameters. The physical mass of both solutions
is set equal to 1. The celestial coordinates $(\alpha,\beta)$ are
measured in the units of physical mass. } }
        \label{WS_a0}
\end{figure}

\begin{figure}[h]
        \setlength{\tabcolsep}{ 0 pt }{\scriptsize\tt
        \begin{tabular}{ cccc }
            \includegraphics[width=4.1cm]{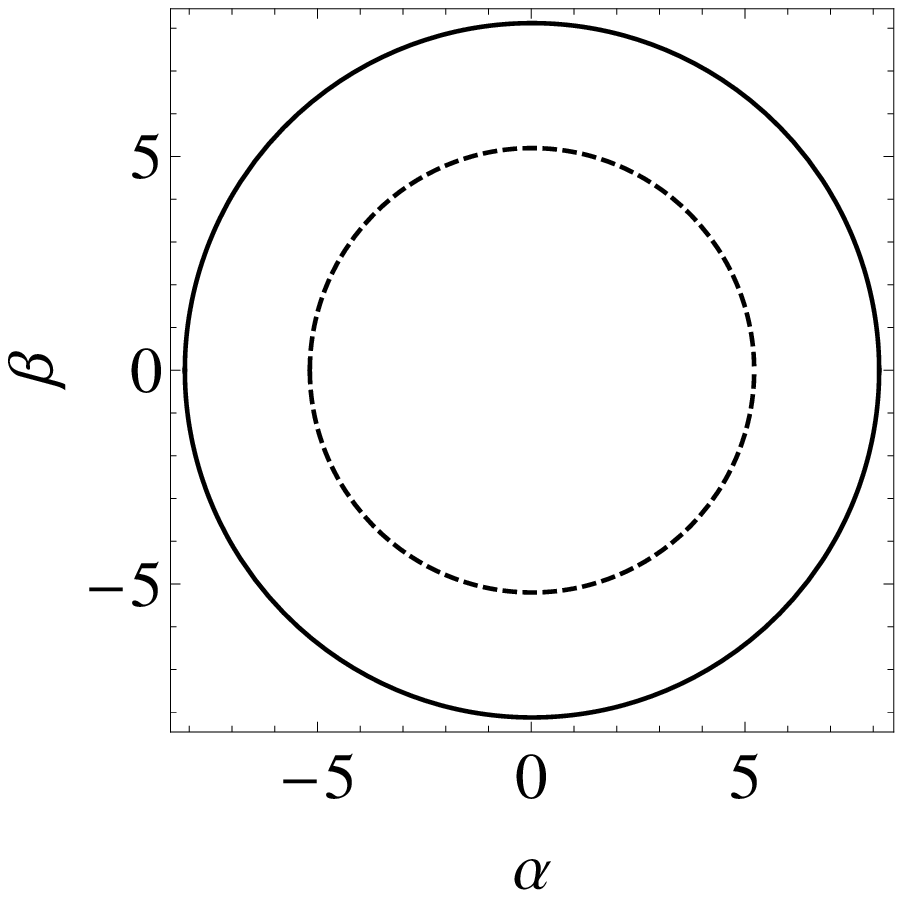} &
            \includegraphics[width=4.1cm]{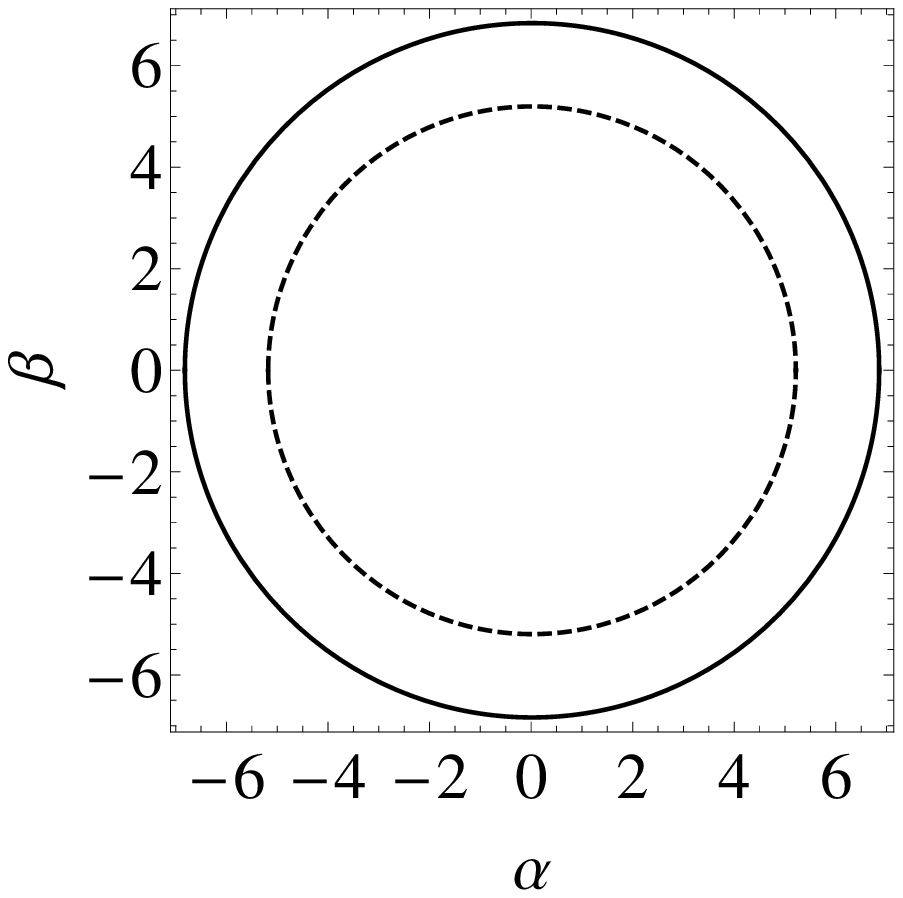} &
            \includegraphics[width=4.1cm]{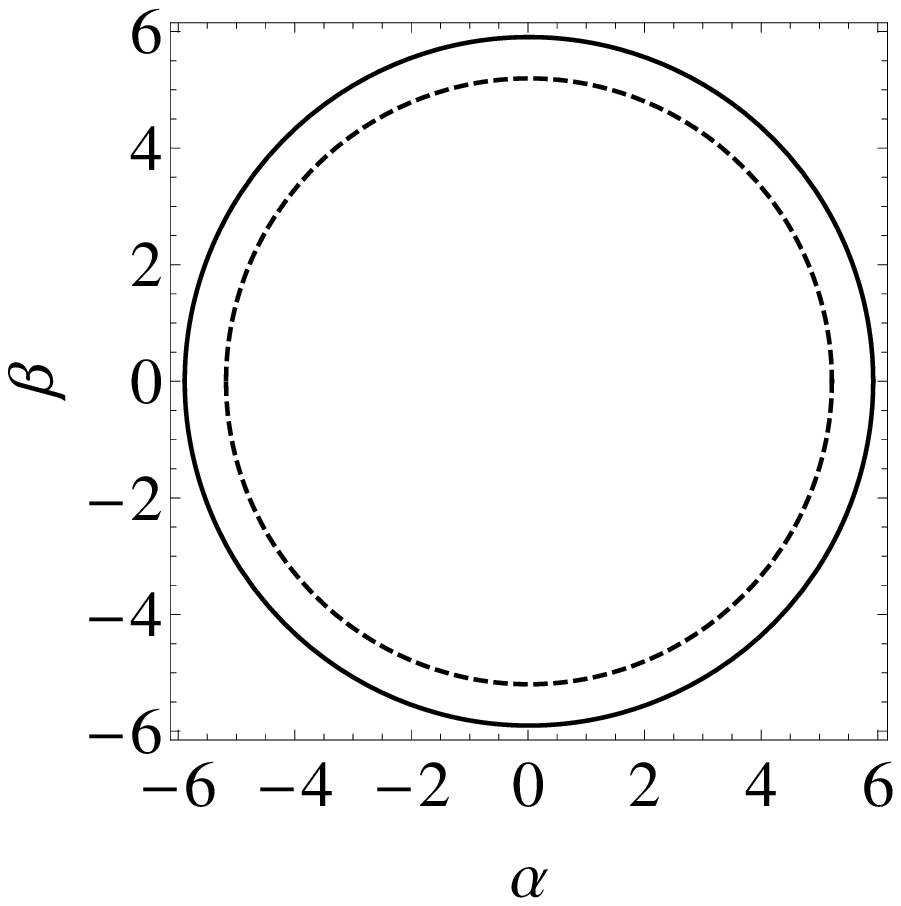} &
            \includegraphics[width=4.1cm]{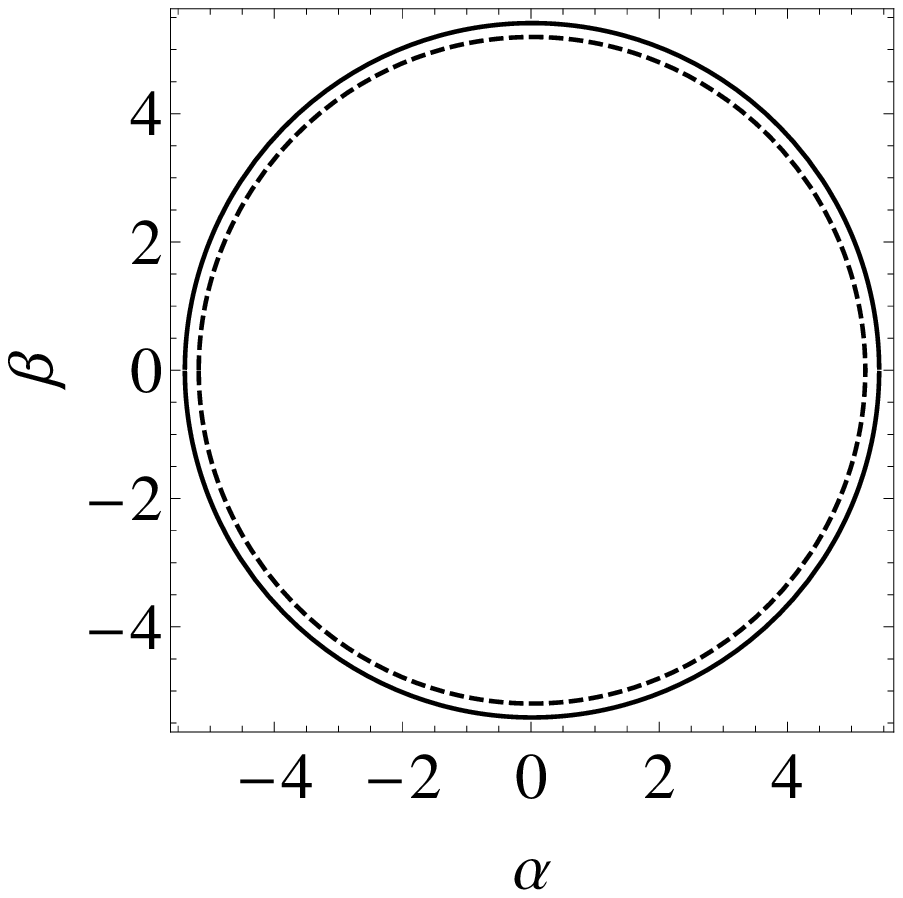} \\
            $a/M_{phys}=0.01$, $\zeta=0.64$;\  &
            $a/M_{phys}=0.01$, $\zeta=0.76$;\  &
            $a/M_{phys}=0.01$, $\zeta=0.88$;\  &
            $a/M_{phys}=0.01$, $\zeta=0.96$ \\
            \includegraphics[width=4.1cm]{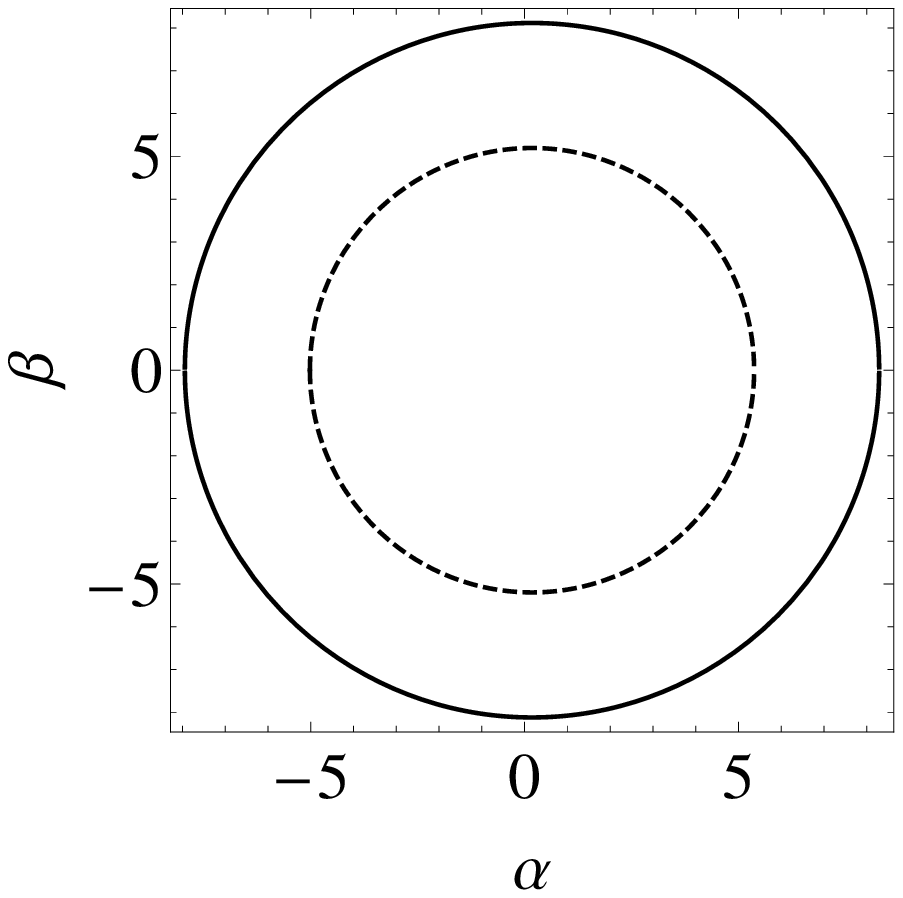} &
            \includegraphics[width=4.1cm]{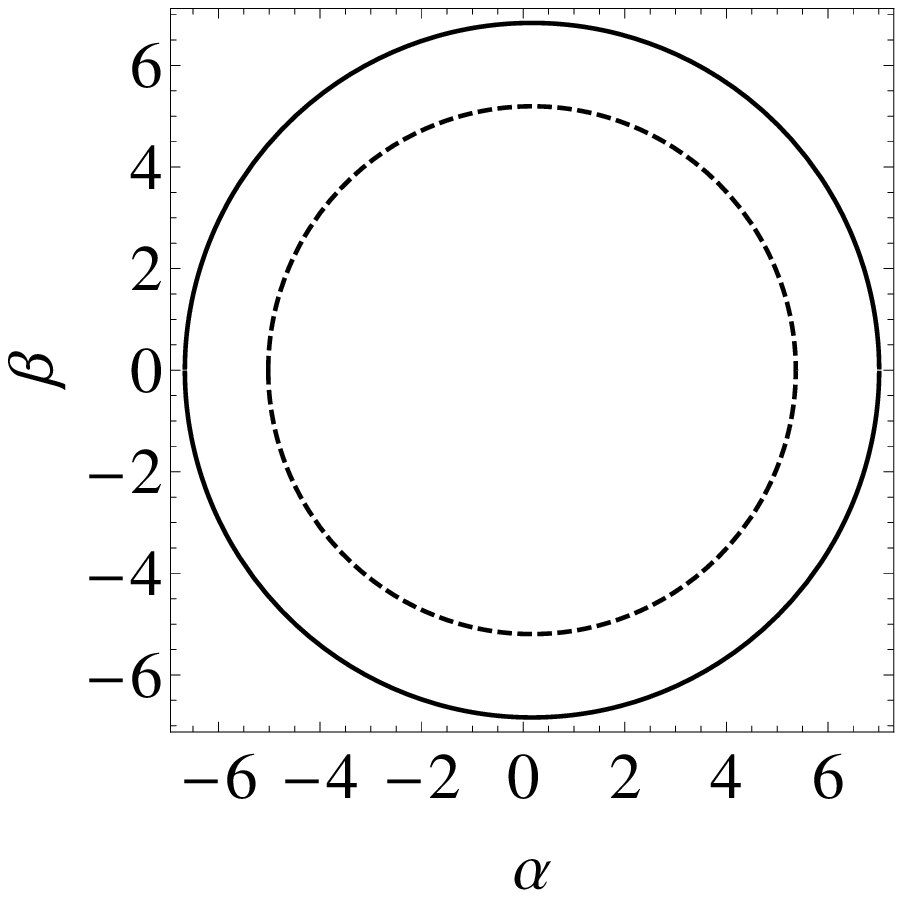} &
            \includegraphics[width=4.1cm]{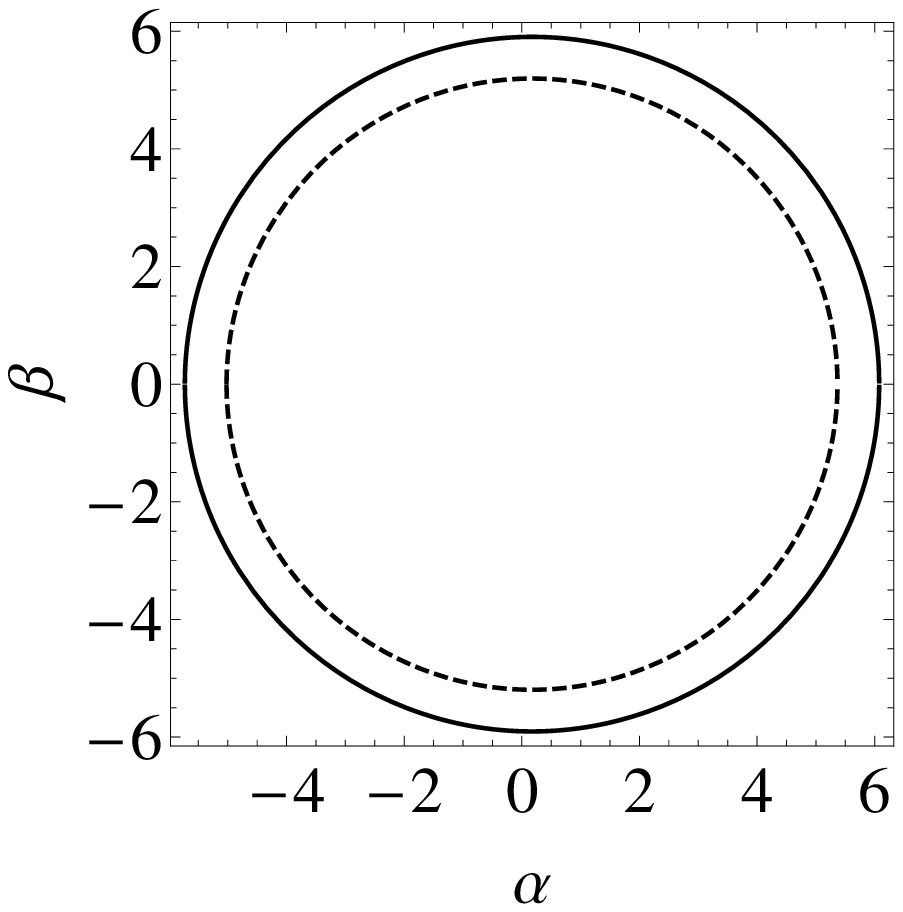} &
            \includegraphics[width=4.1cm]{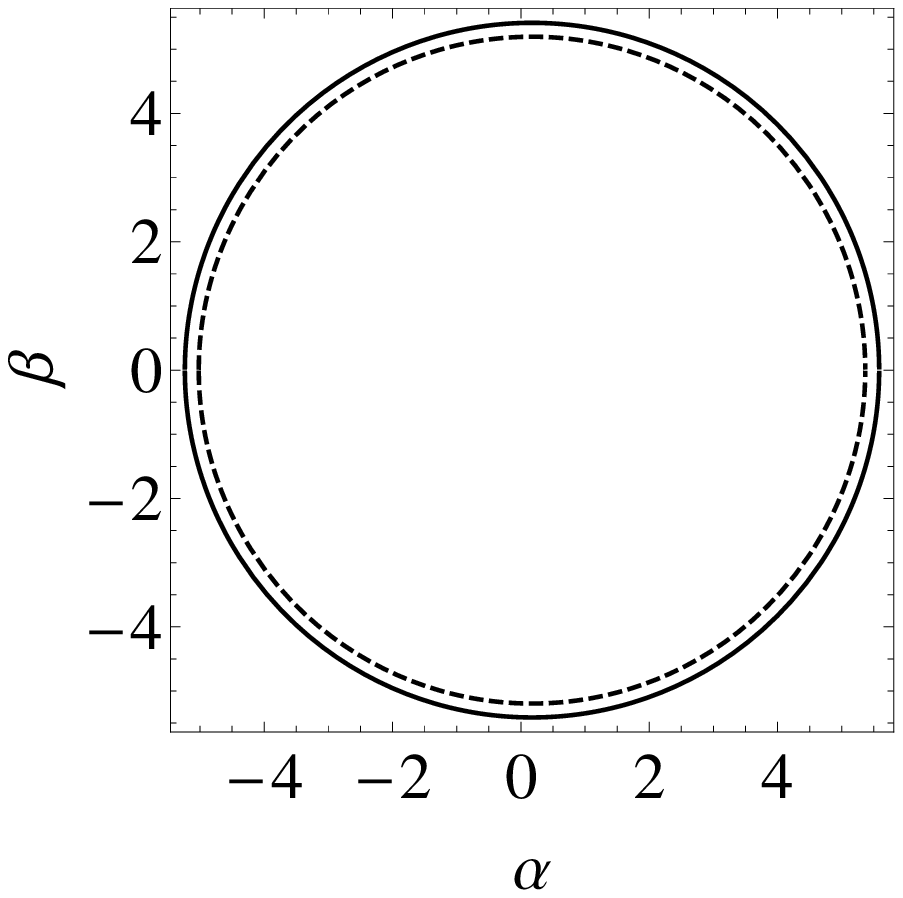} \\
            $a/M_{phys}=0.1$, $\zeta=0.64$;\  &
            $a/M_{phys}=0.1$, $\zeta=0.76$;\  &
            $a/M_{phys}=0.1$, $\zeta=0.88$;\  &
            $a/M_{phys}=0.1$, $\zeta=0.96$ \\
            \includegraphics[width=4.1cm]{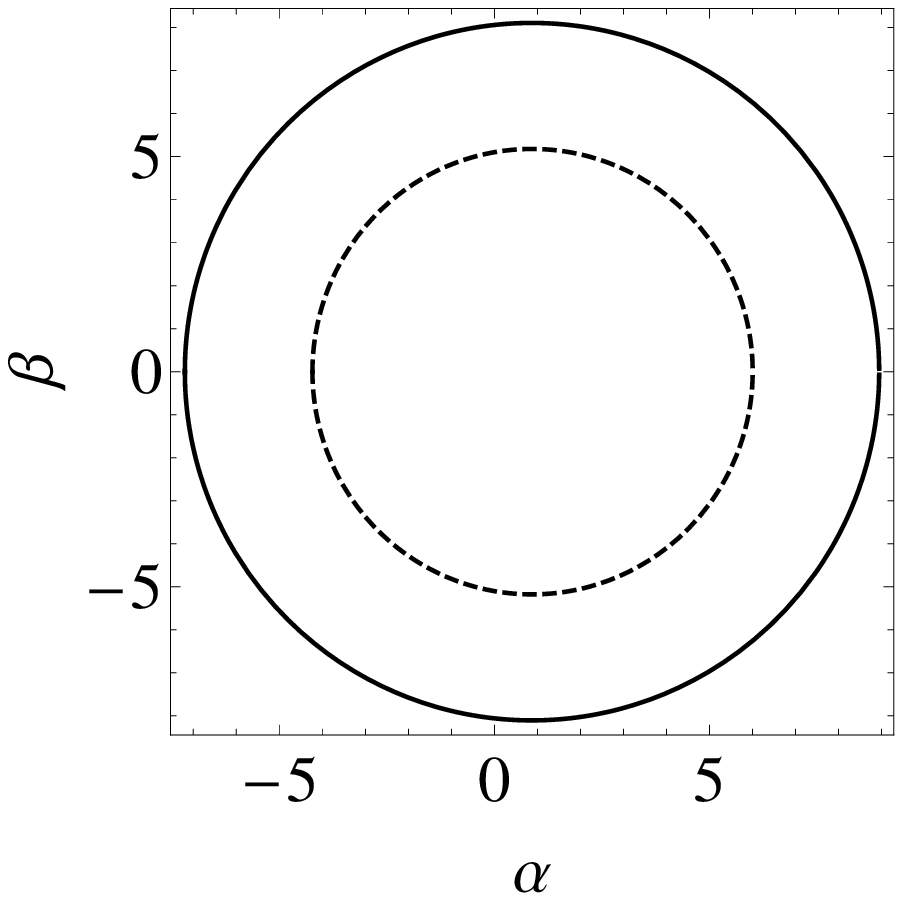} &
            \includegraphics[width=4.1cm]{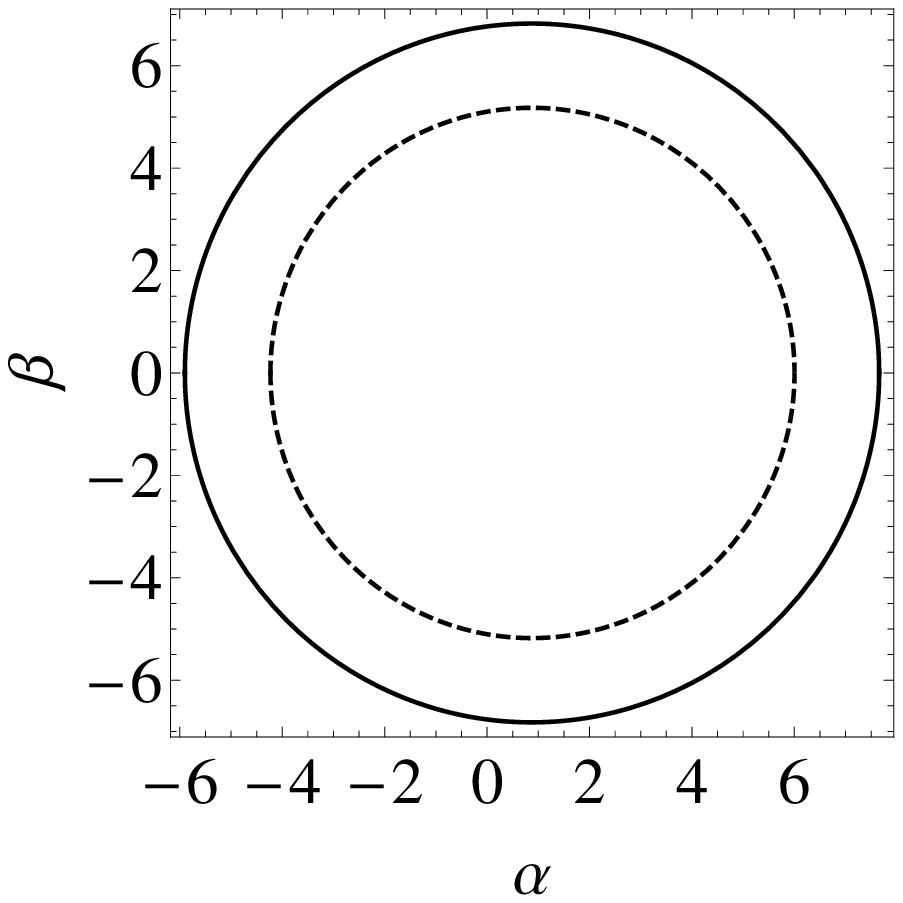} &
            \includegraphics[width=4.1cm]{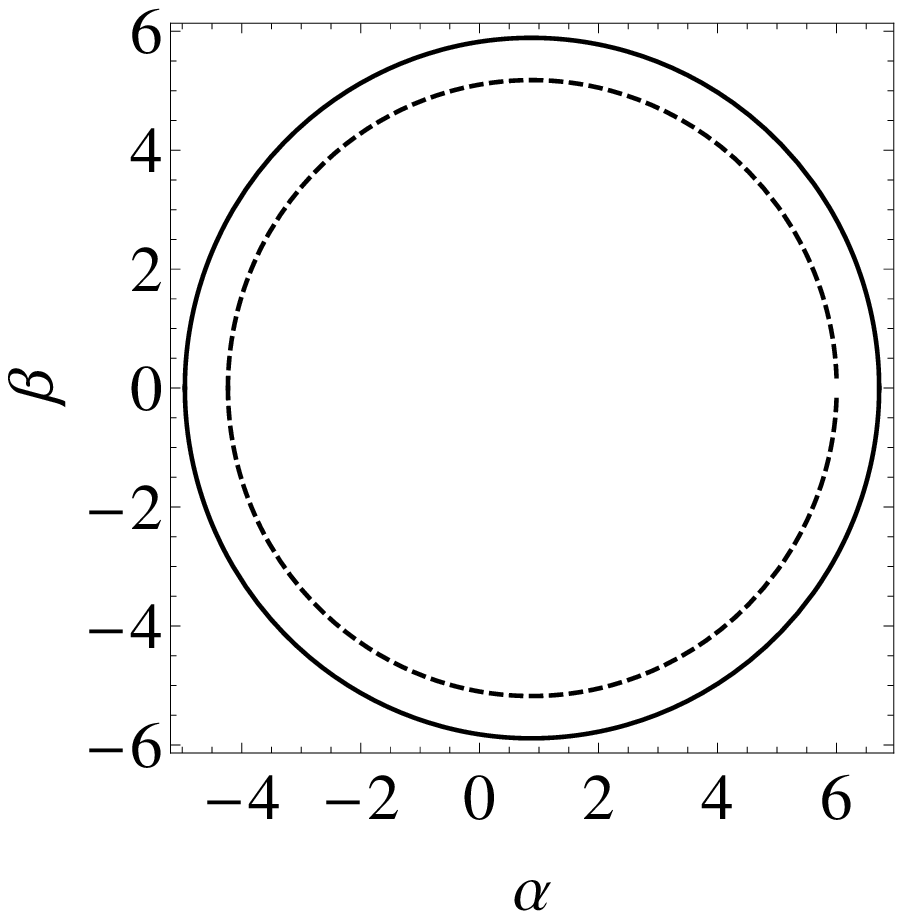} &
            \includegraphics[width=4.1cm]{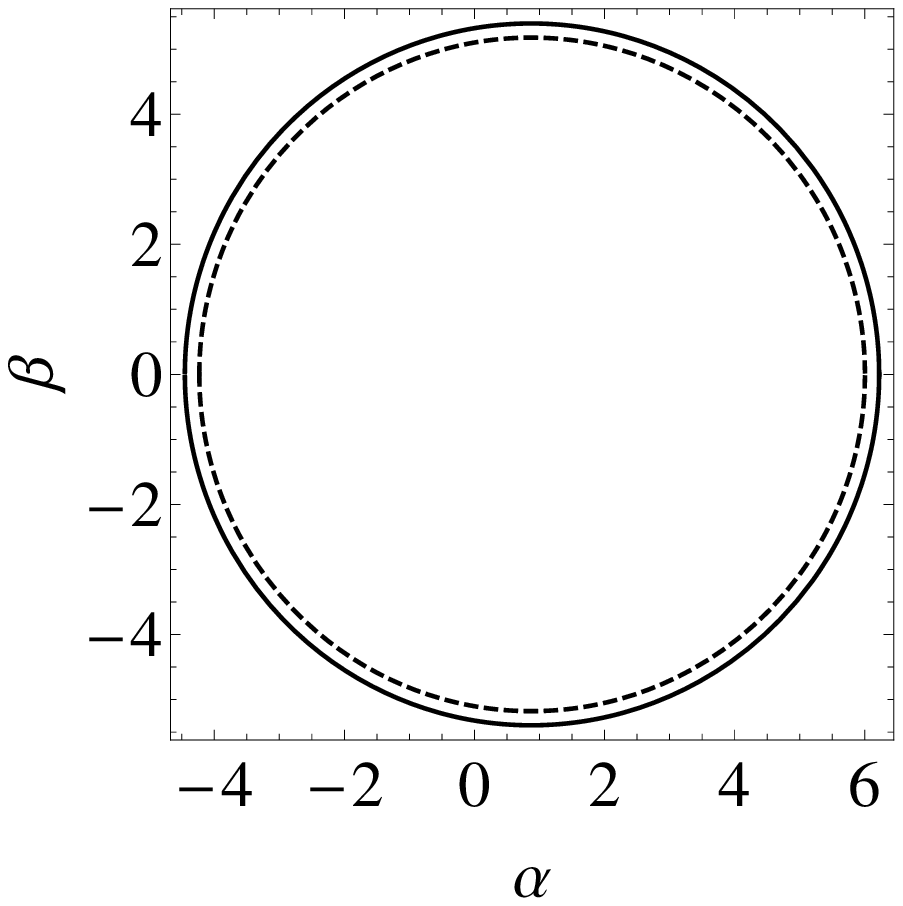} \\
            $a/M_{phys}=0.5$, $\zeta=0.64$;\  &
            $a/M_{phys}=0.5$, $\zeta=0.76$;\  &
            $a/M_{phys}=0.5$, $\zeta=0.88$;\  &
            $a/M_{phys}=0.5$, $\zeta=0.96$ \\
            \includegraphics[width=4.1cm]{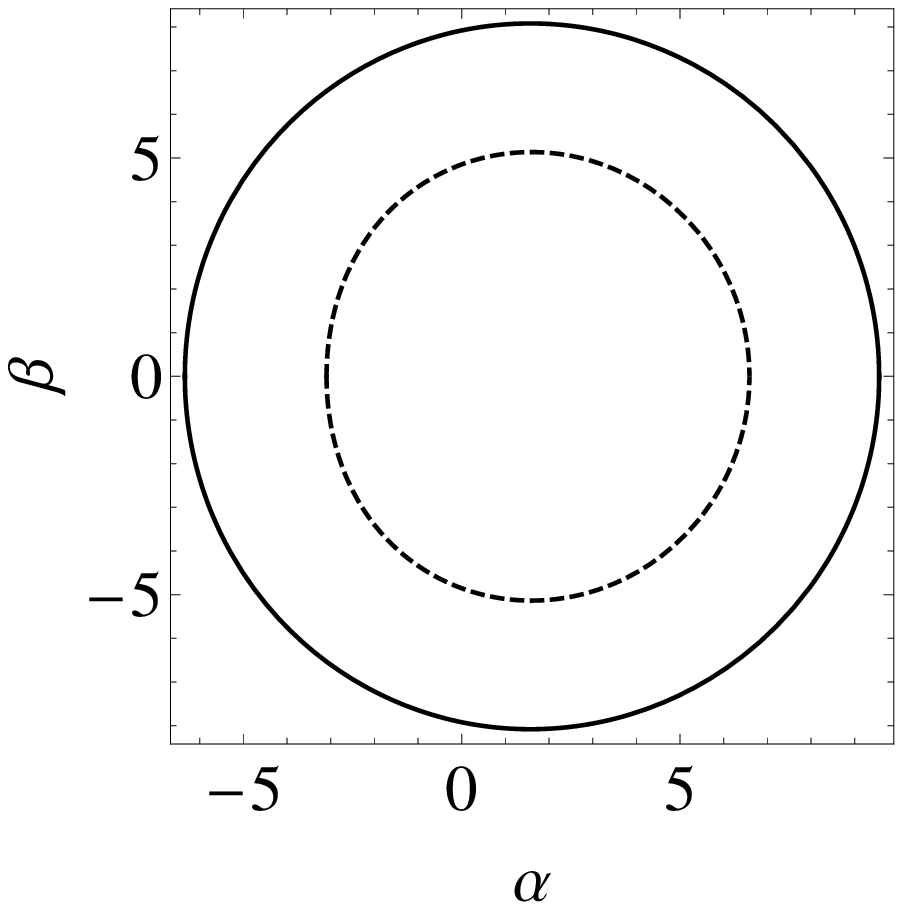} &
            \includegraphics[width=4.1cm]{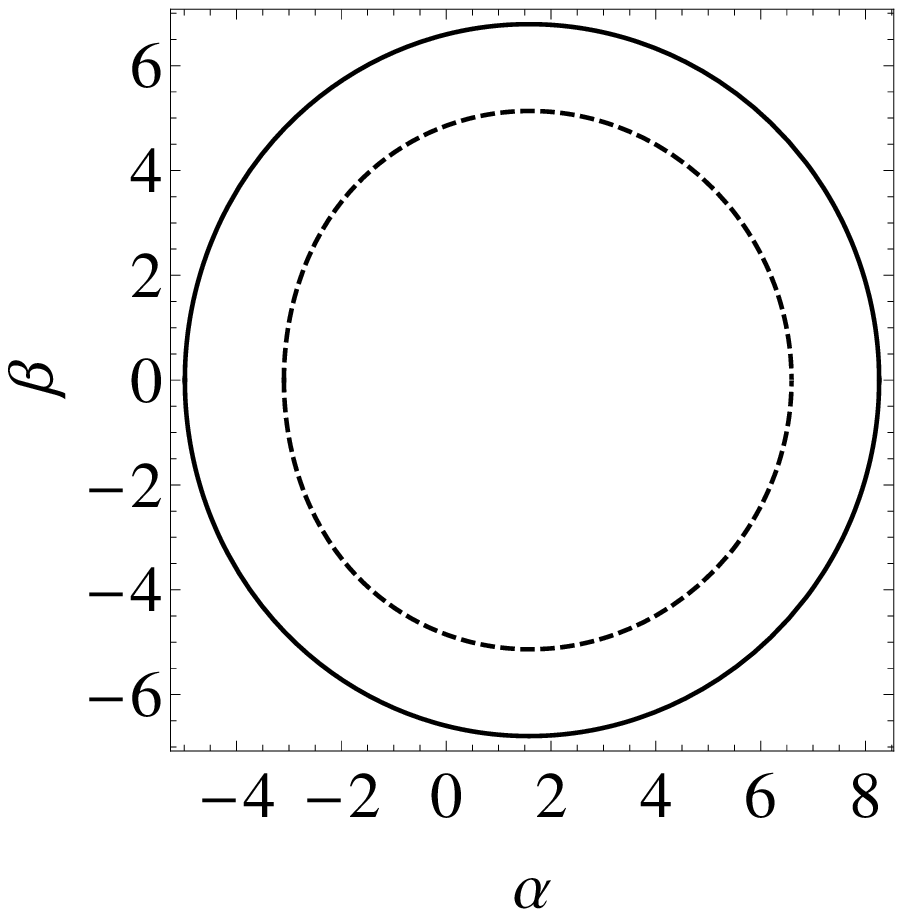} &
            \includegraphics[width=4.1cm]{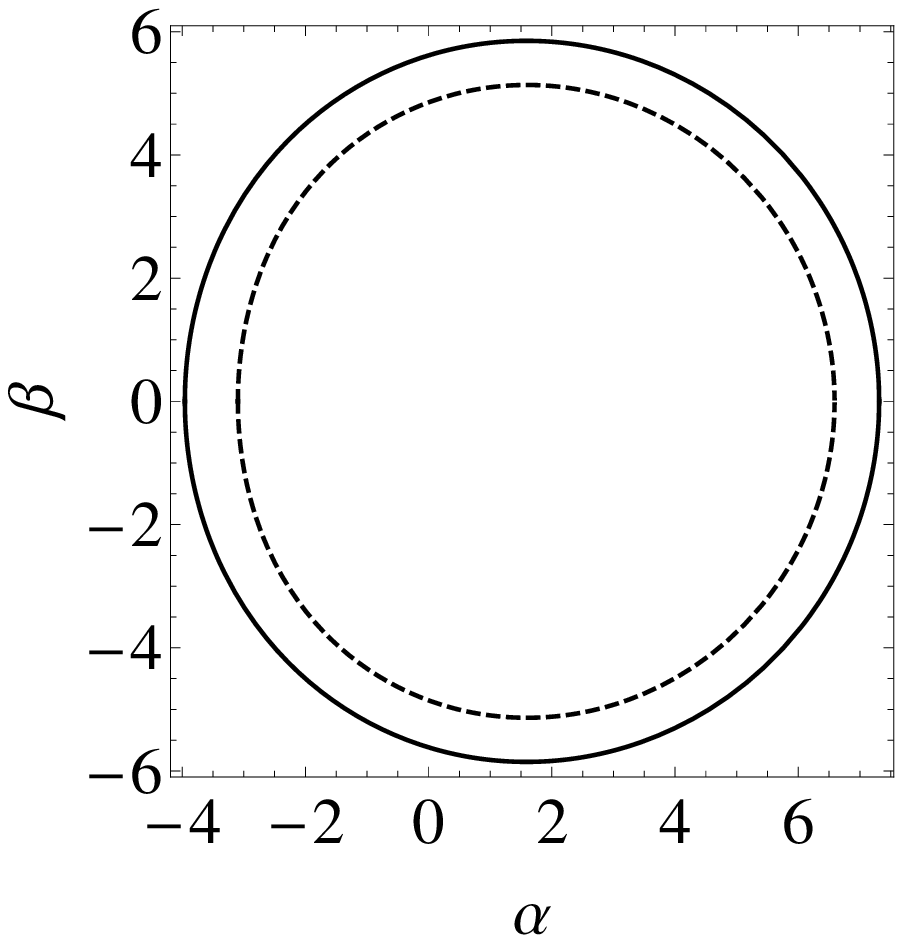} &
            \includegraphics[width=4.1cm]{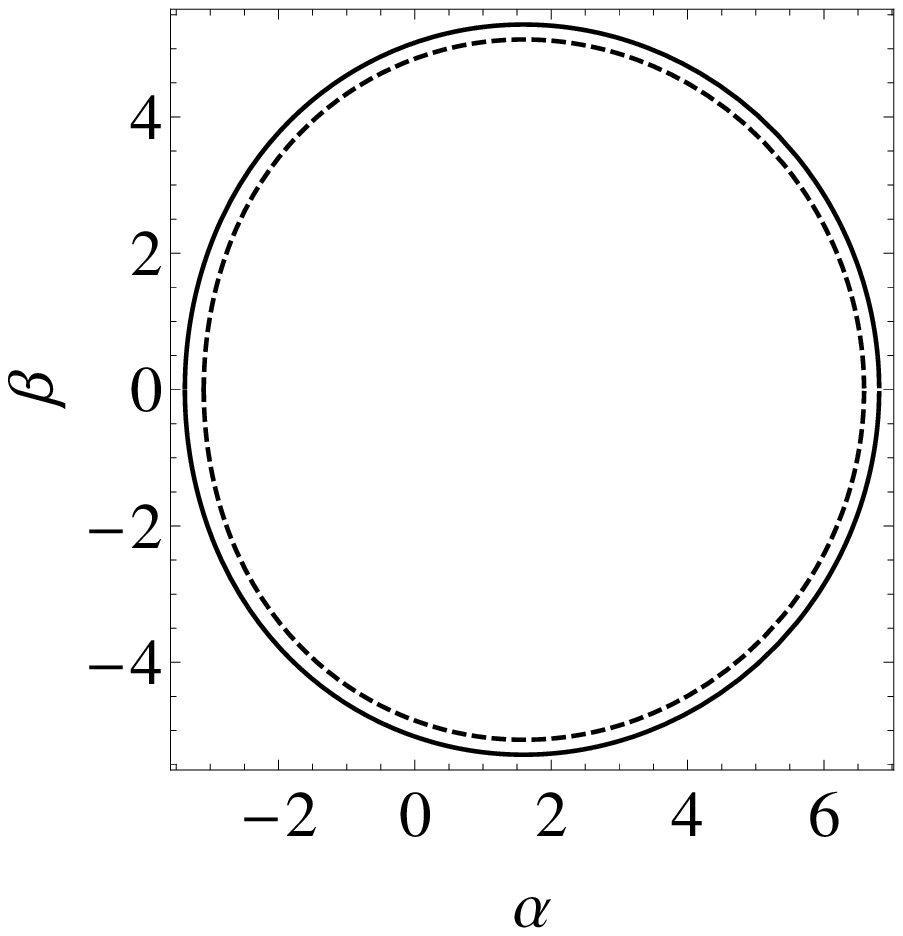} \\
            $a/M_{phys}=0.9$, $\zeta=0.64$;\  &
            $a/M_{phys}=0.9$, $\zeta=0.76$;\  &
            $a/M_{phys}=0.9$, $\zeta=0.88$;\  &
            $a/M_{phys}=0.9$, $\zeta=0.96$ \\
        \end{tabular}}
\caption{\footnotesize{The shadow of the Kerr black hole pierced by
a cosmic string (solid line) and the Kerr black hole (dashed line)
with inclination angle $\theta_{0}=\pi/3\ rad$ for different
rotation and string parameters. The physical mass of both solutions
is set equal to 1. The celestial coordinates $(\alpha,\beta)$ are
measured in the units of physical mass. } }
        \label{WS_a1}
\end{figure}

\begin{figure}[h]
        \setlength{\tabcolsep}{ 0 pt }{\scriptsize\tt
        \begin{tabular}{ cccc }
            \includegraphics[width=4.1cm]{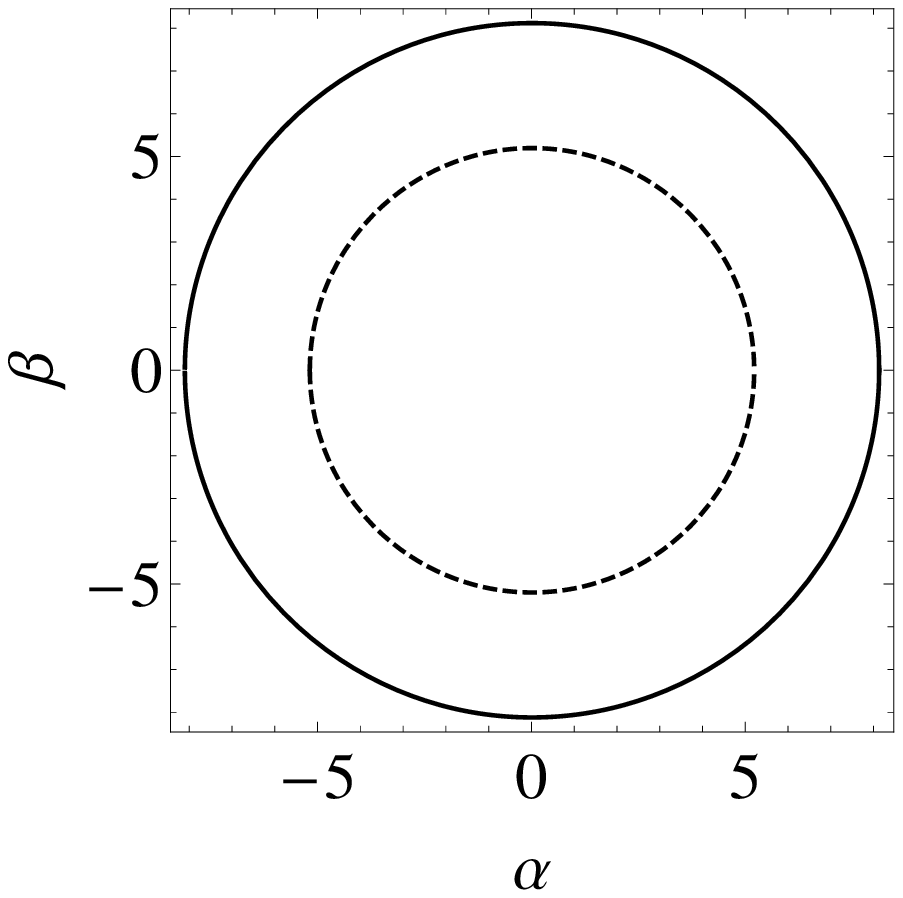} &
            \includegraphics[width=4.1cm]{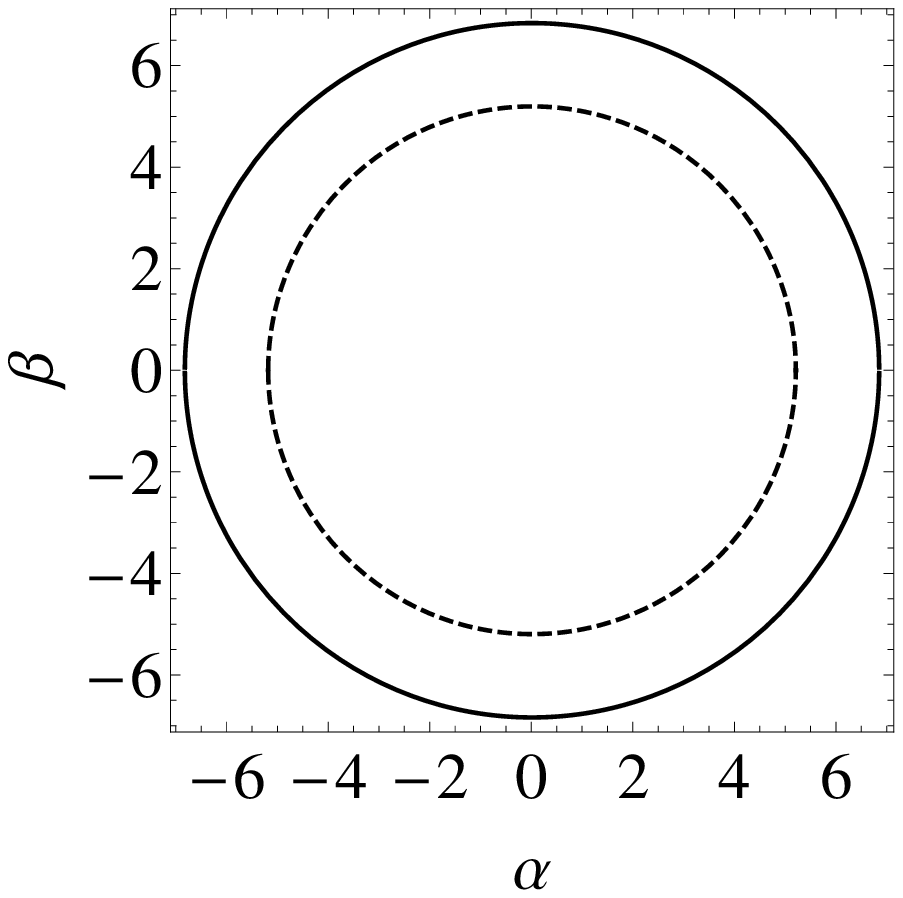} &
            \includegraphics[width=4.1cm]{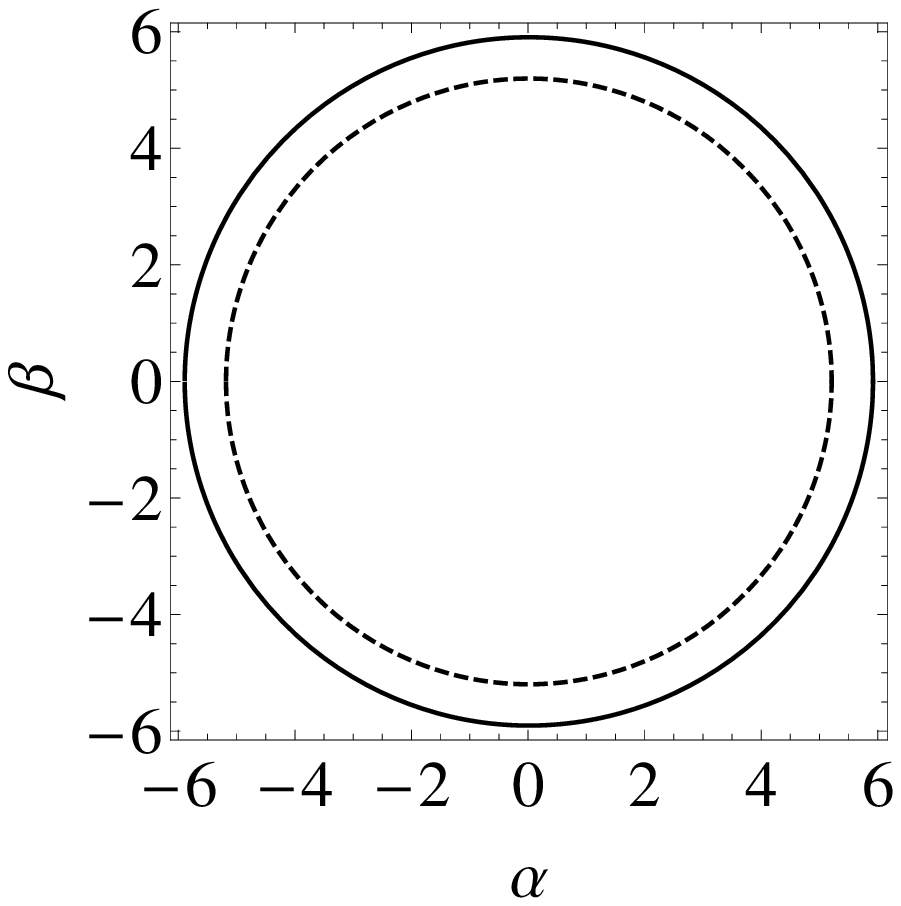} &
            \includegraphics[width=4.1cm]{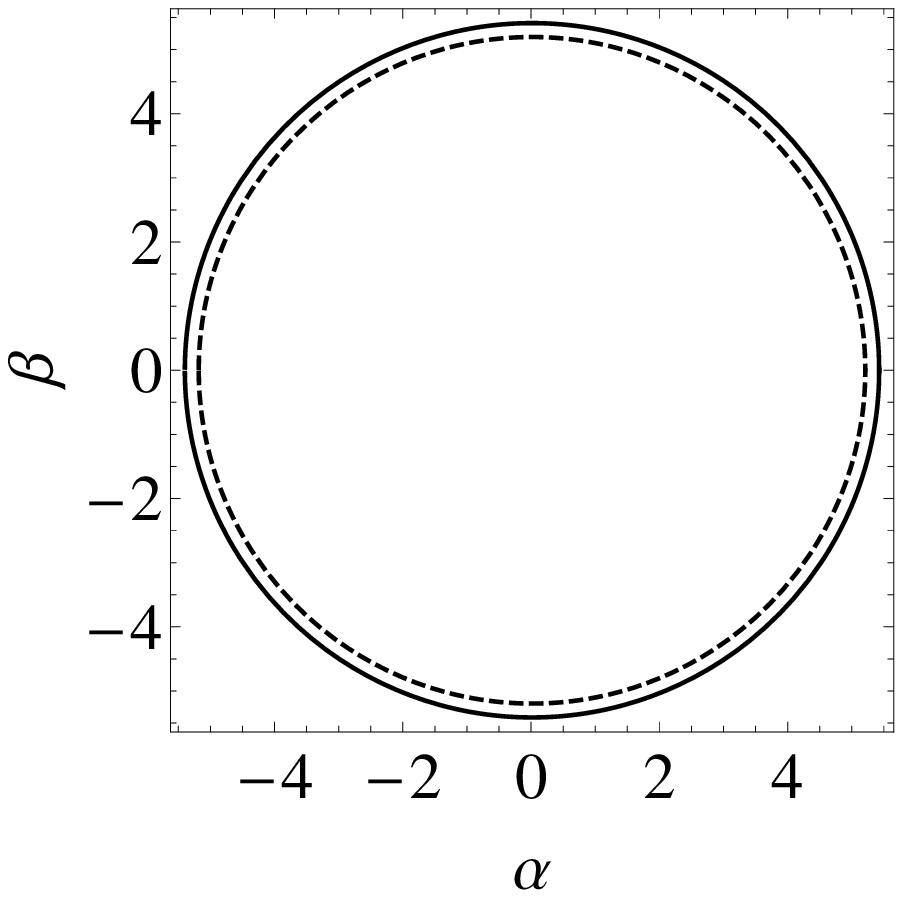} \\
            $a/M_{phys}=0.01$, $\zeta=0.64$;\  &
            $a/M_{phys}=0.01$, $\zeta=0.76$;\  &
            $a/M_{phys}=0.01$, $\zeta=0.88$;\  &
            $a/M_{phys}=0.01$, $\zeta=0.96$ \\
            \includegraphics[width=4.1cm]{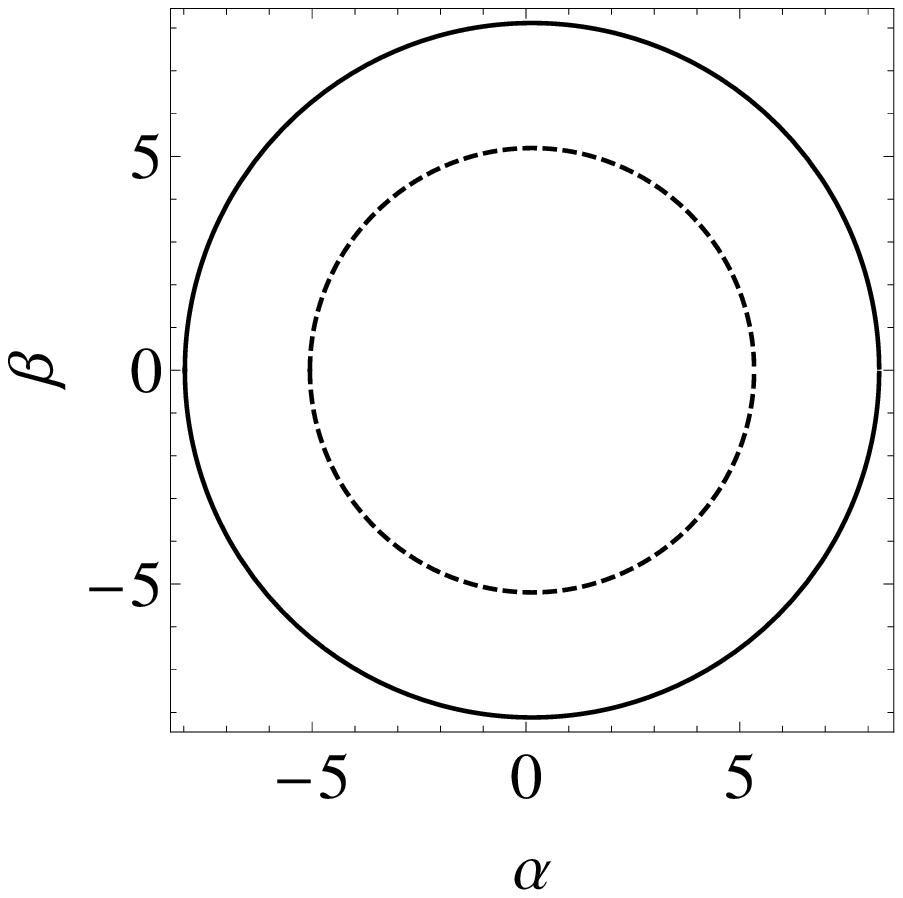} &
            \includegraphics[width=4.1cm]{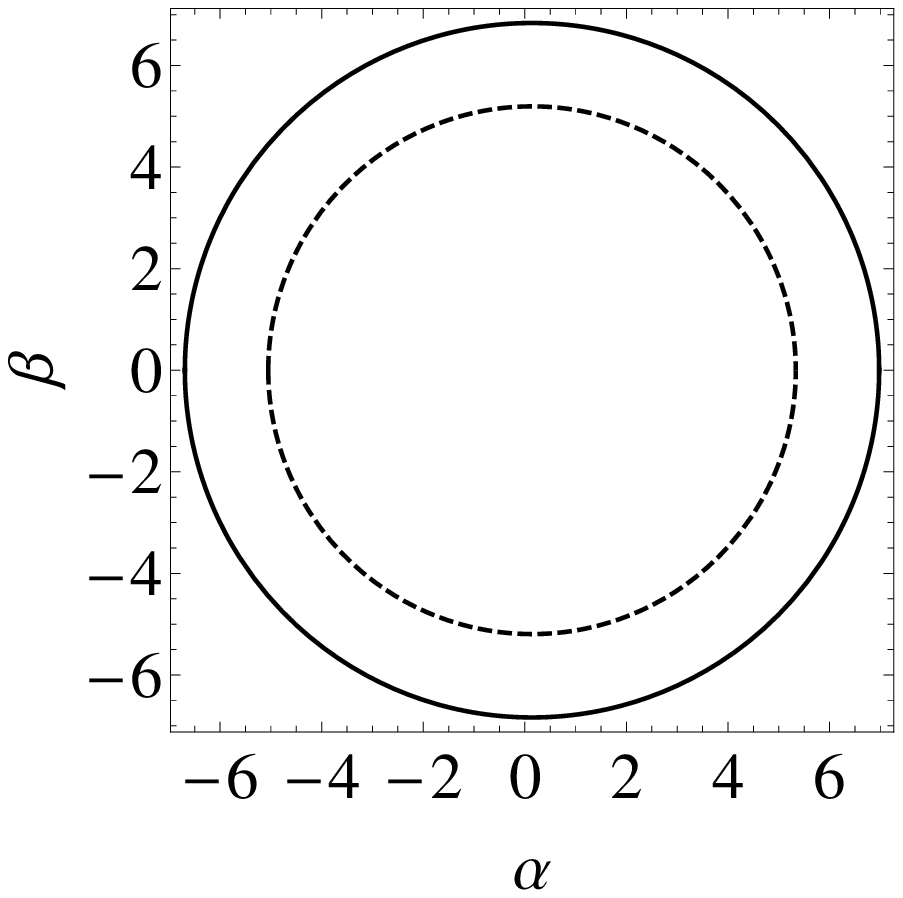} &
            \includegraphics[width=4.1cm]{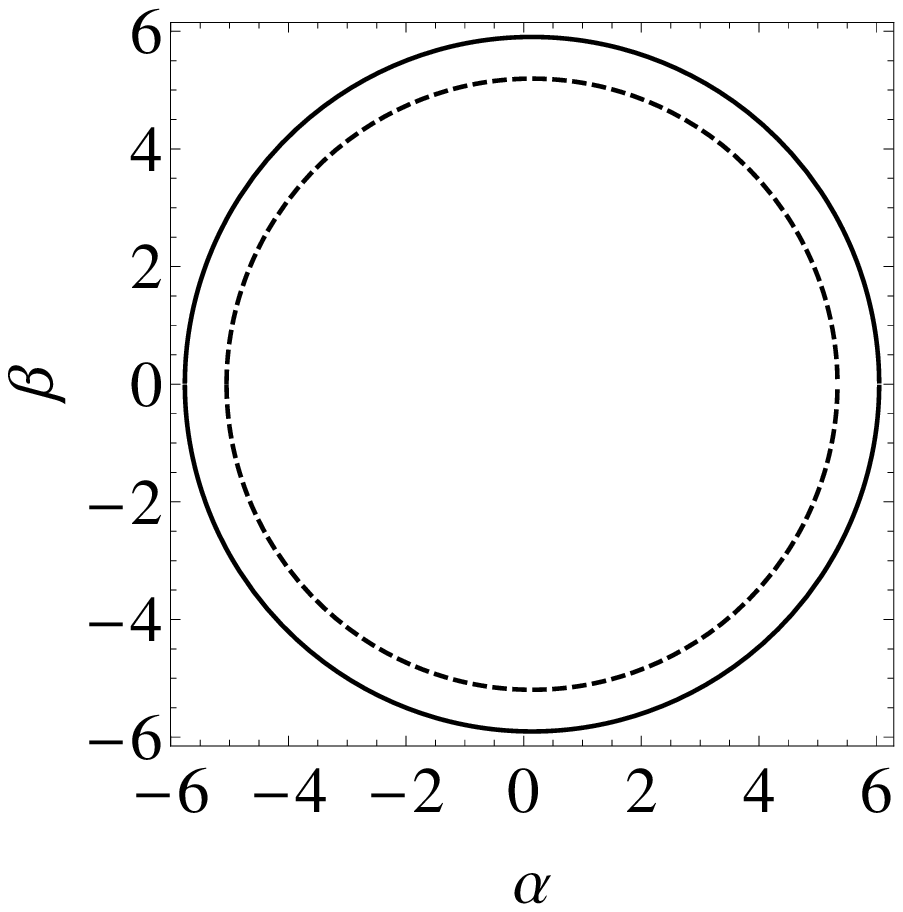} &
            \includegraphics[width=4.1cm]{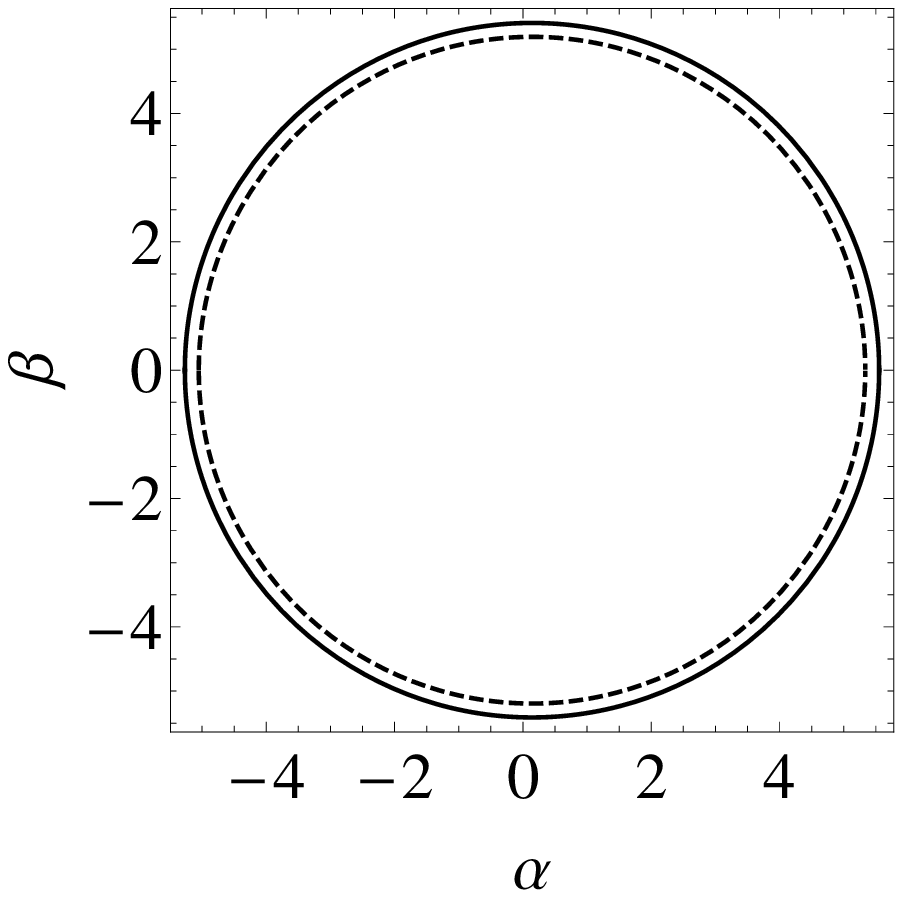} \\
            $a/M_{phys}=0.1$, $\zeta=0.64$;\  &
            $a/M_{phys}=0.1$, $\zeta=0.76$;\  &
            $a/M_{phys}=0.1$, $\zeta=0.88$;\  &
            $a/M_{phys}=0.1$, $\zeta=0.96$ \\
            \includegraphics[width=4.1cm]{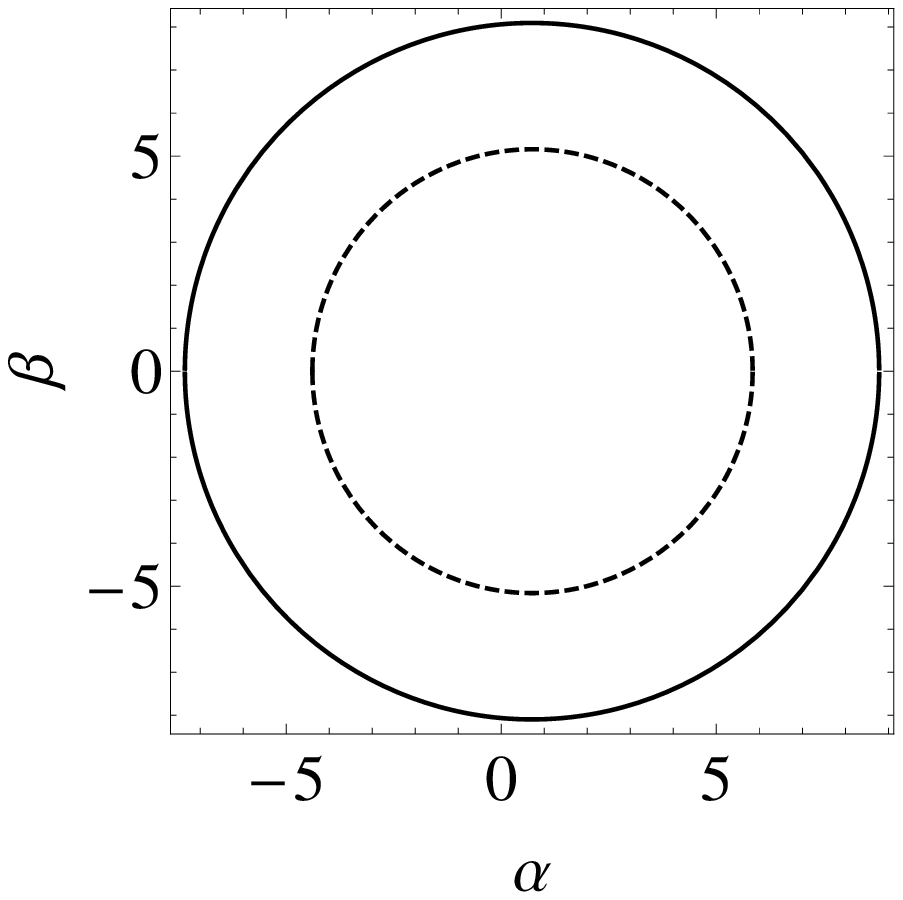} &
            \includegraphics[width=4.1cm]{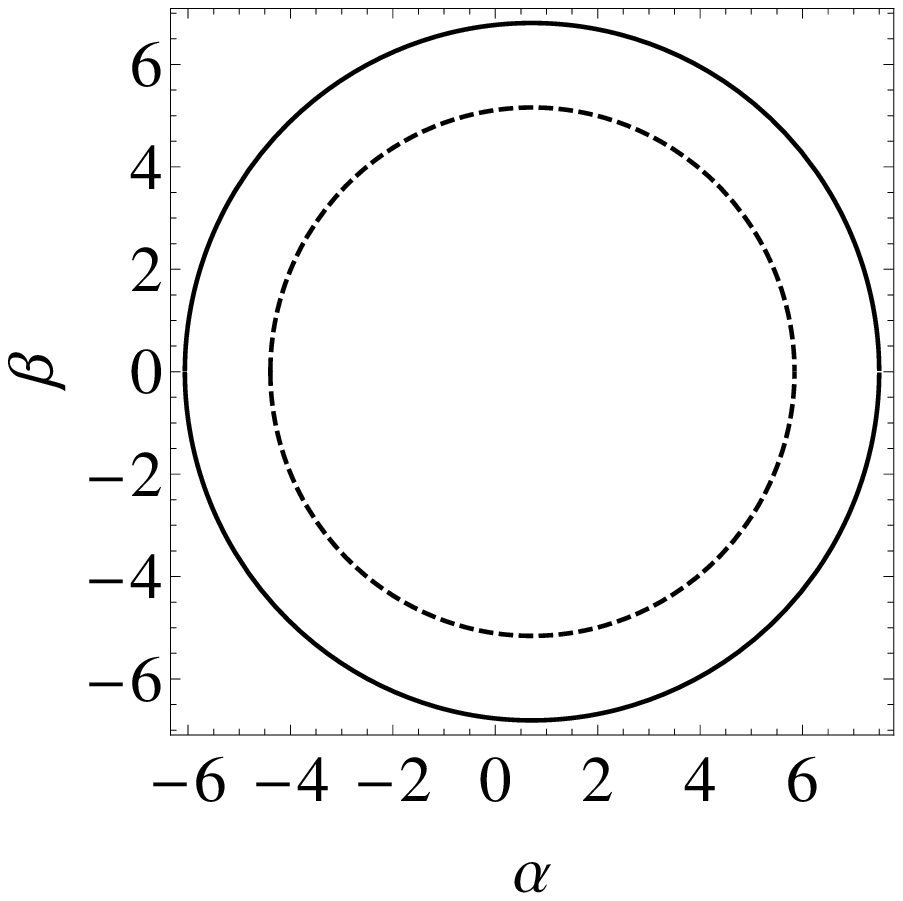} &
            \includegraphics[width=4.1cm]{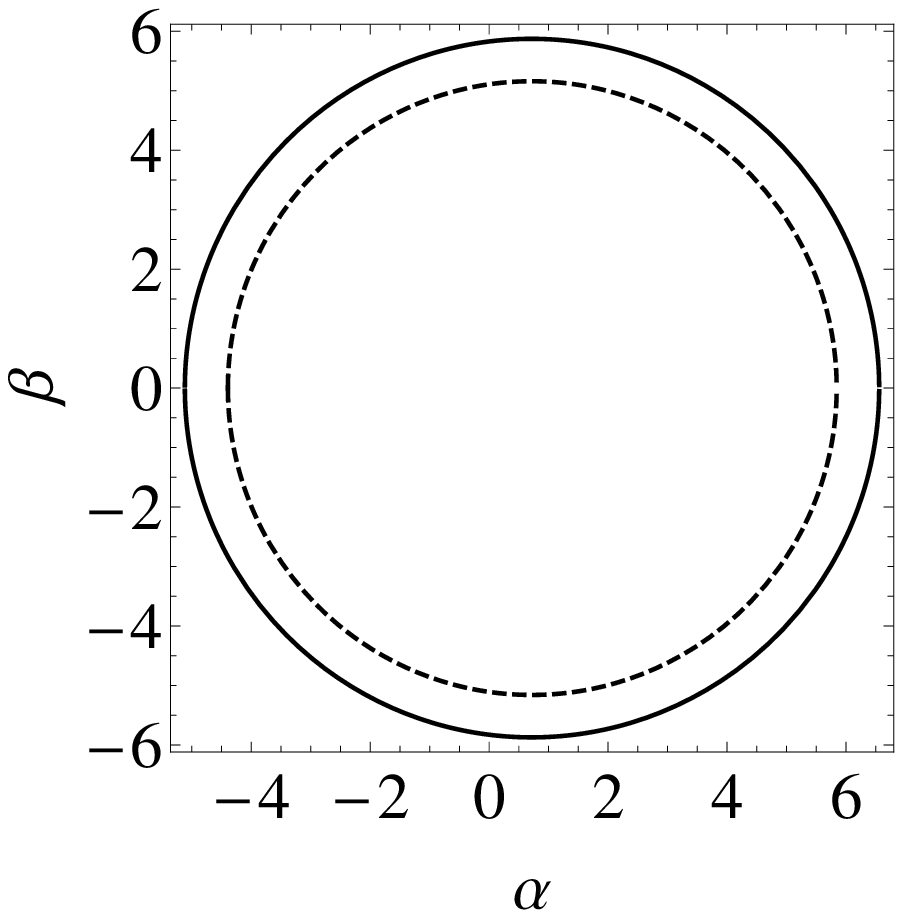} &
            \includegraphics[width=4.1cm]{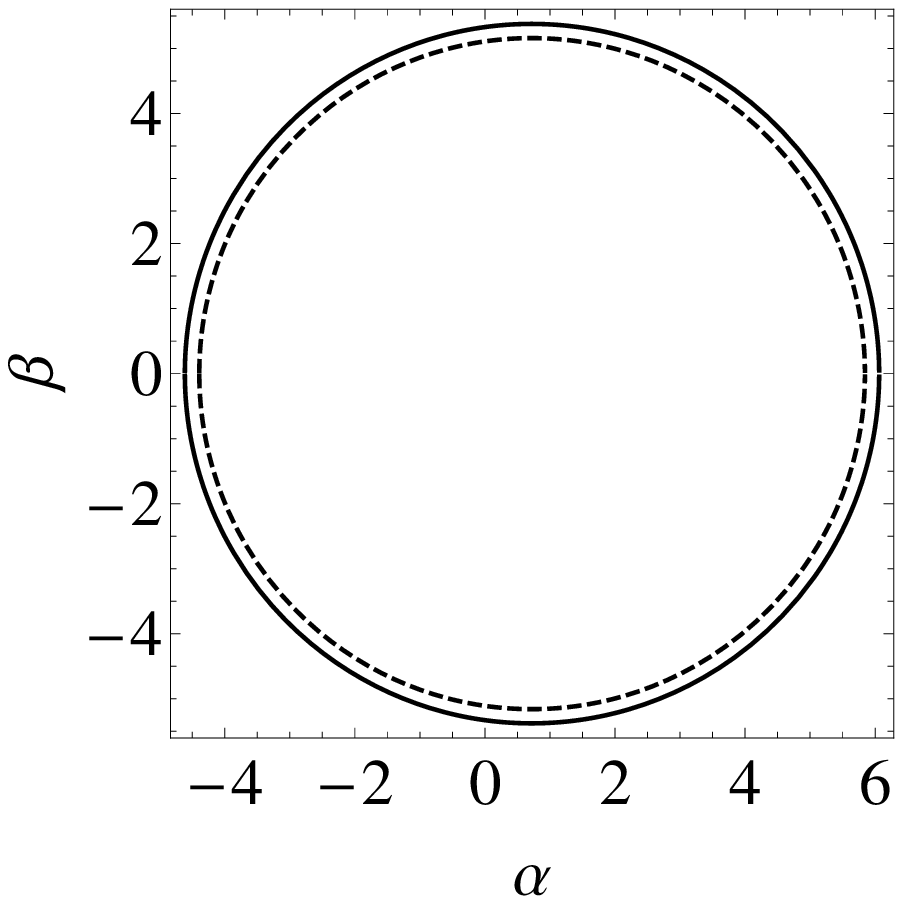} \\
            $a/M_{phys}=0.5$, $\zeta=0.64$;\  &
            $a/M_{phys}=0.5$, $\zeta=0.76$;\  &
            $a/M_{phys}=0.5$, $\zeta=0.88$;\  &
            $a/M_{phys}=0.5$, $\zeta=0.96$ \\
            \includegraphics[width=4.1cm]{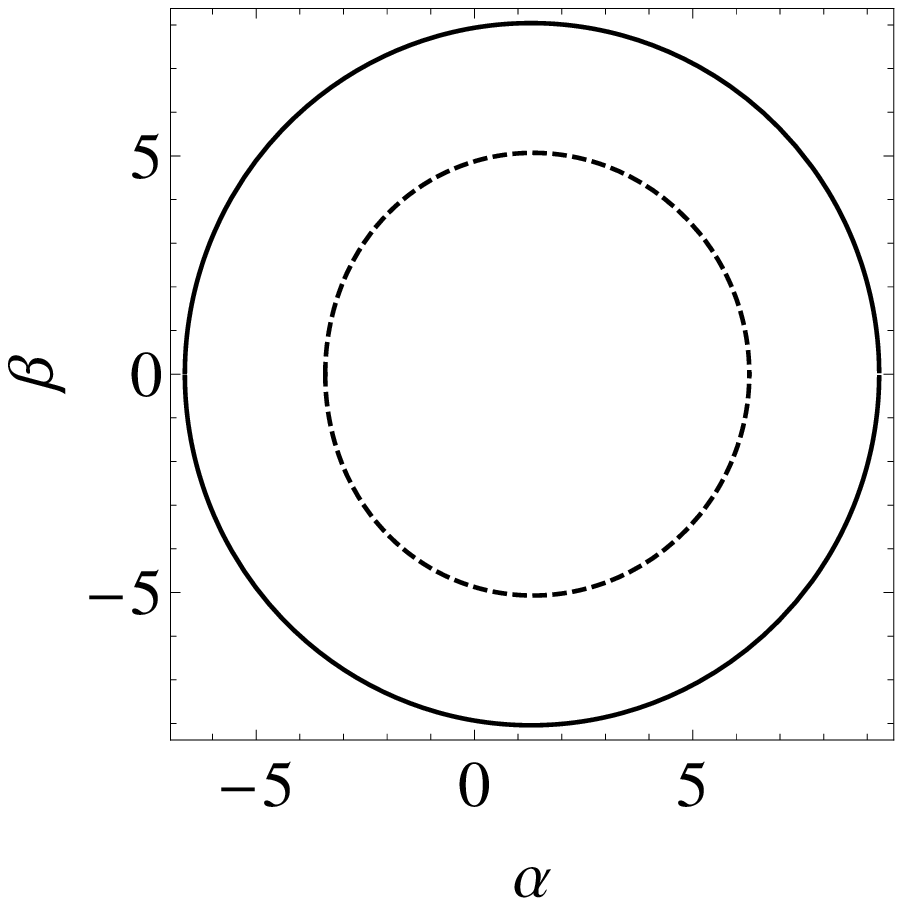} &
            \includegraphics[width=4.1cm]{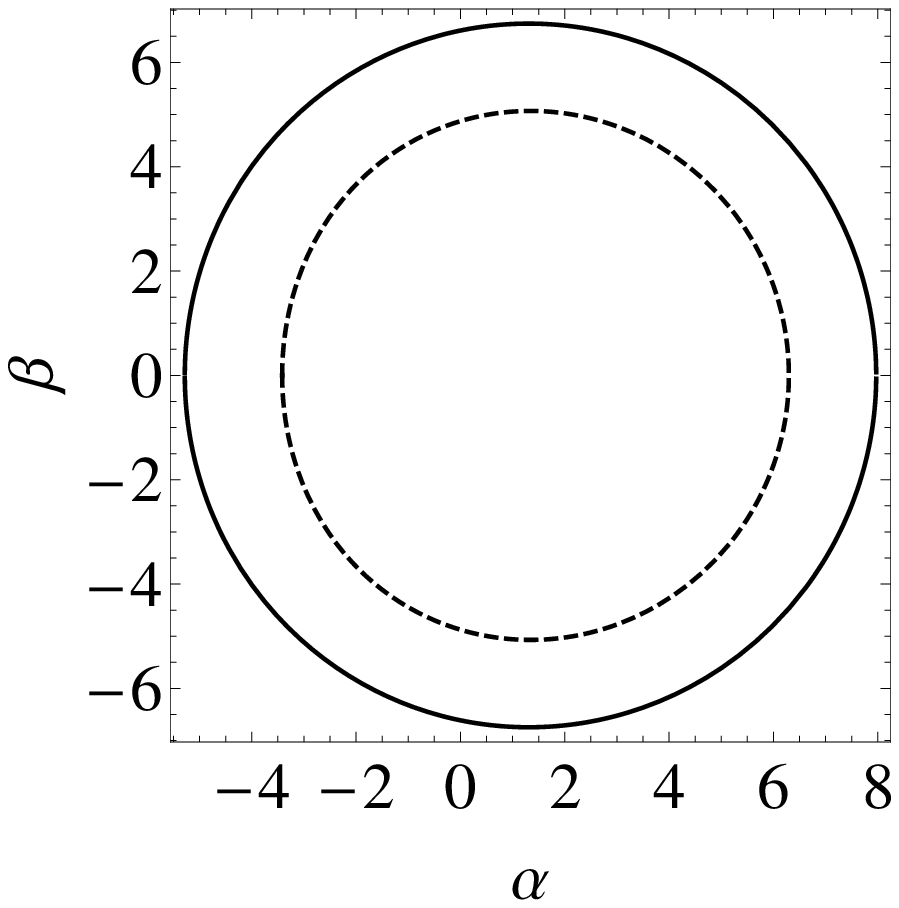} &
            \includegraphics[width=4.1cm]{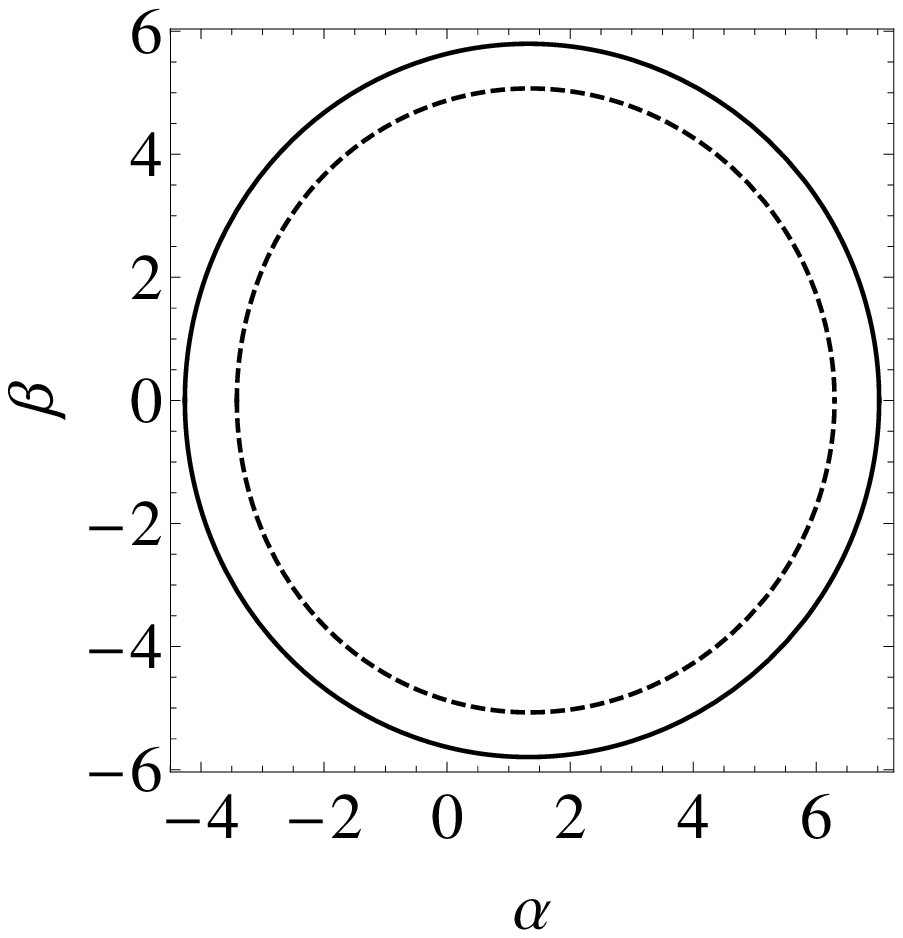} &
            \includegraphics[width=4.1cm]{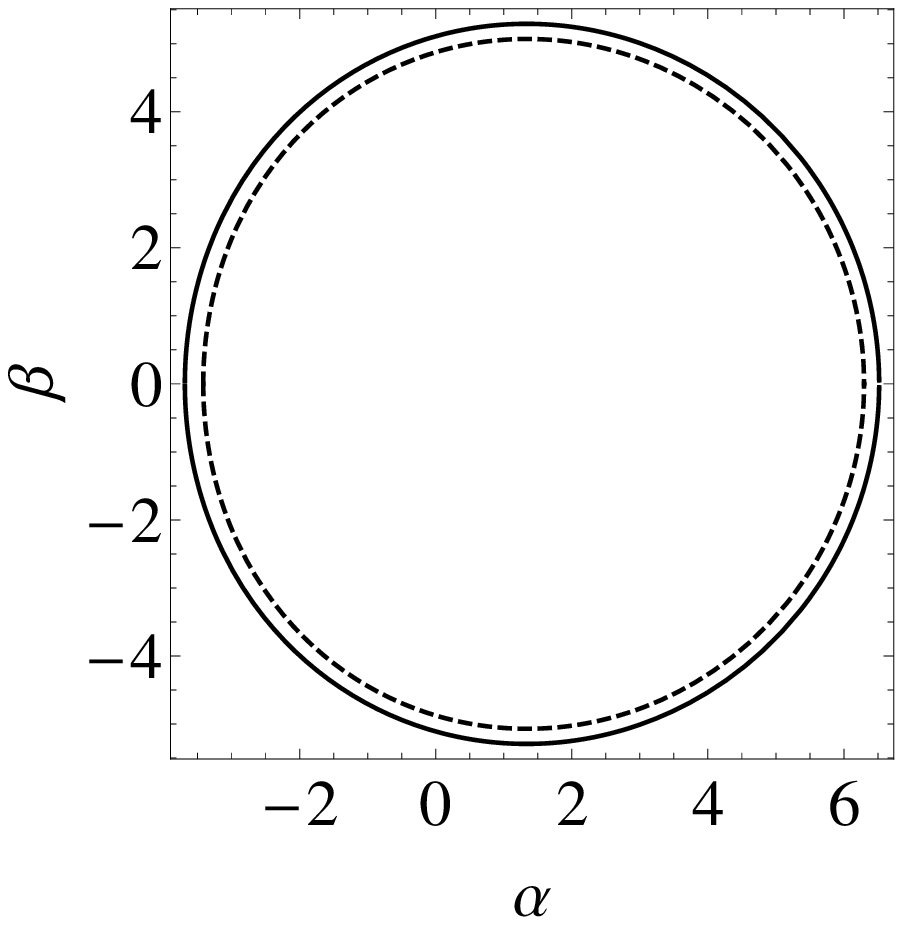} \\
            $a/M_{phys}=0.9$, $\zeta=0.64$;\  &
            $a/M_{phys}=0.9$, $\zeta=0.76$;\  &
            $a/M_{phys}=0.9$, $\zeta=0.88$;\  &
            $a/M_{phys}=0.9$, $\zeta=0.96$ \\
        \end{tabular}}
\caption{\footnotesize{The shadow of the Kerr black hole pierced by
a cosmic string (solid line) and the Kerr black hole (dashed line)
with inclination angle $\theta_{0}=\pi/4\ rad$ for different
rotation and string parameters. The physical mass of both solutions
is set equal to 1. The celestial coordinates $(\alpha,\beta)$ are
measured in the units of physical mass. } }
        \label{WS_a2}
\end{figure}

\begin{figure}[h]
        \setlength{\tabcolsep}{ 0 pt }{\scriptsize\tt
        \begin{tabular}{ cccc }
            \includegraphics[width=4.1cm]{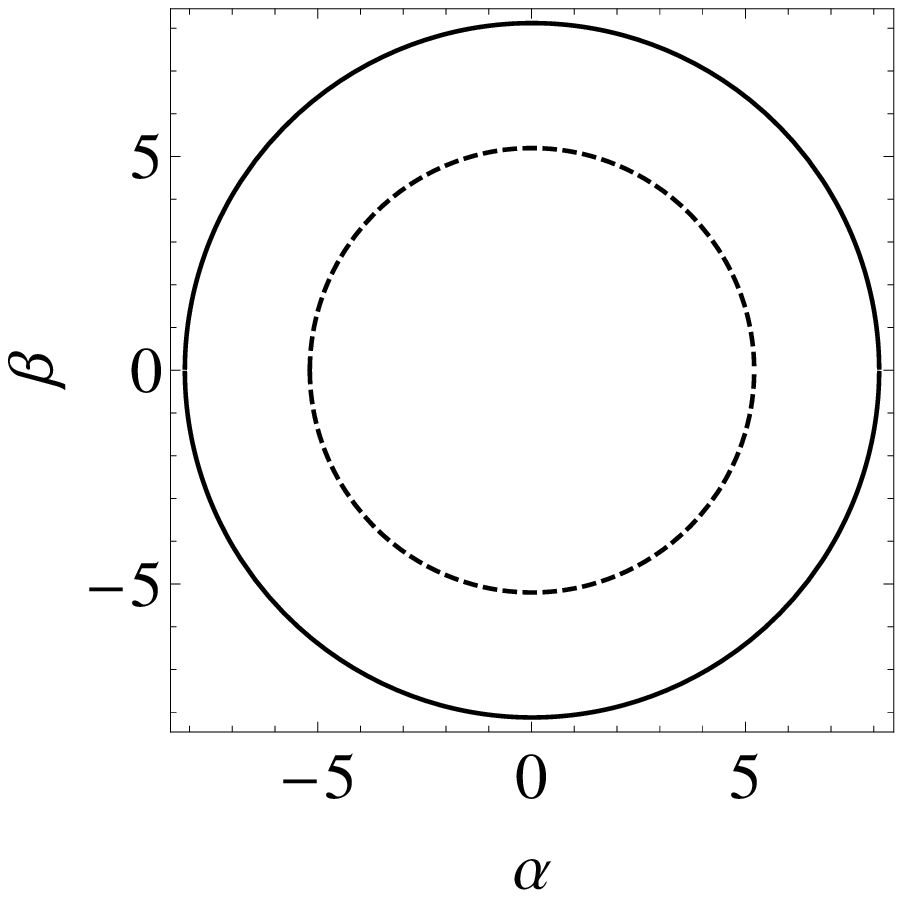} &
            \includegraphics[width=4.1cm]{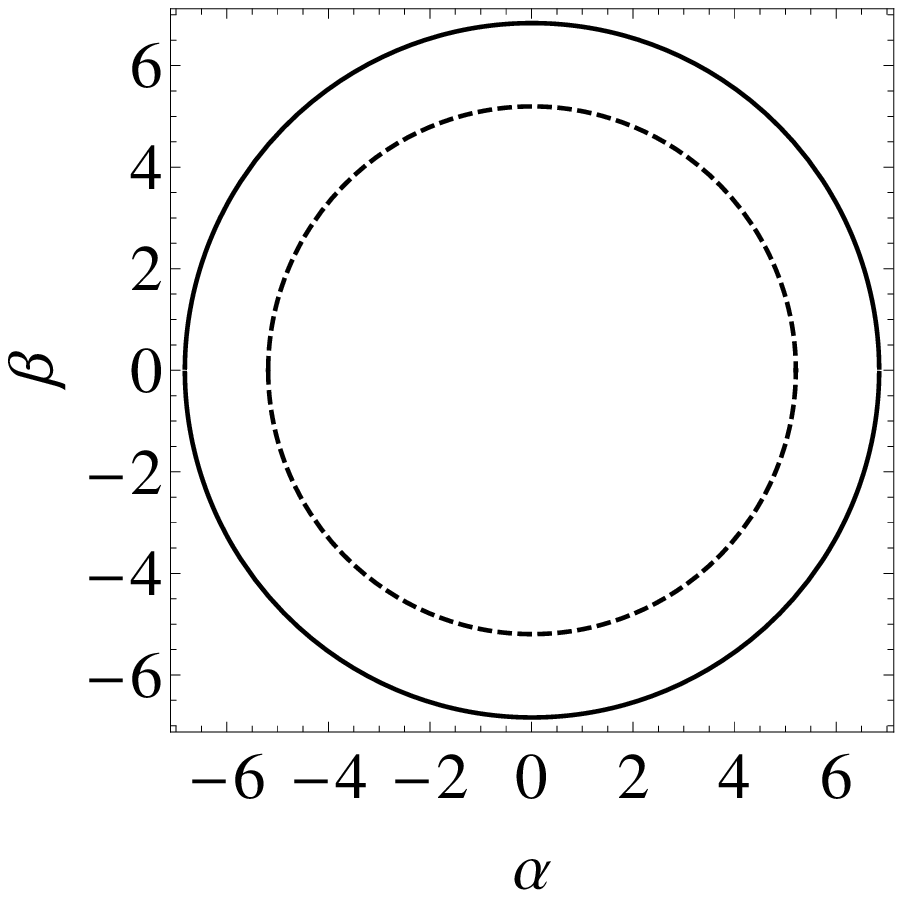} &
            \includegraphics[width=4.1cm]{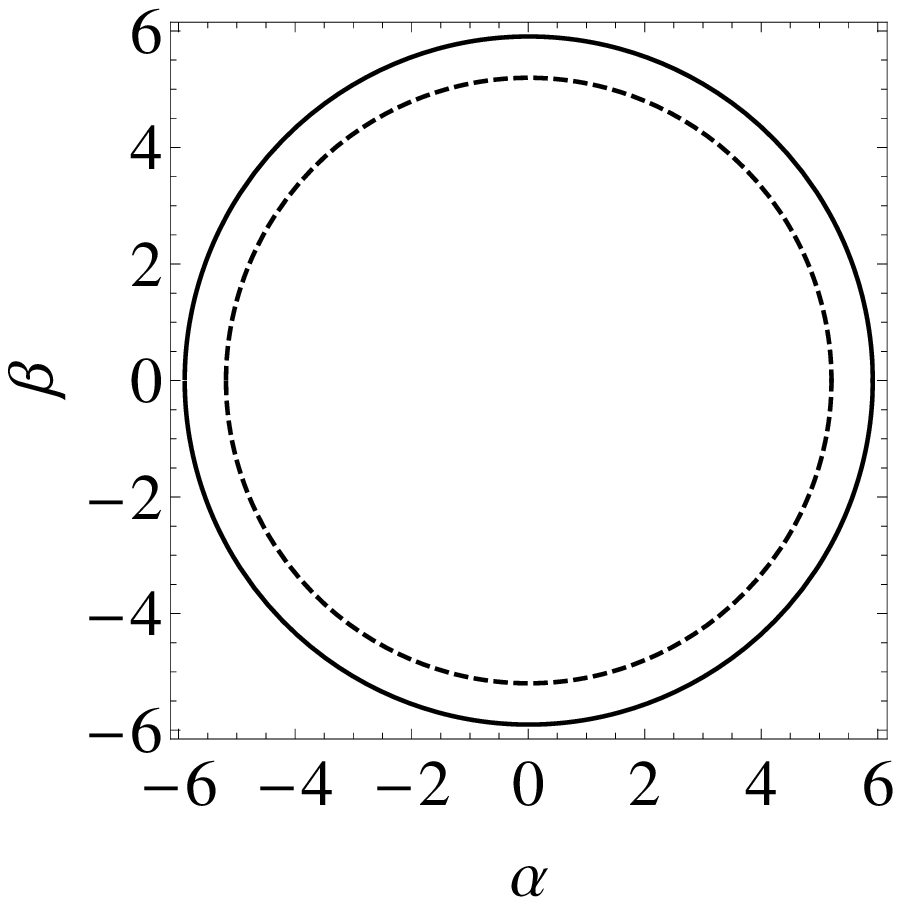} &
            \includegraphics[width=4.1cm]{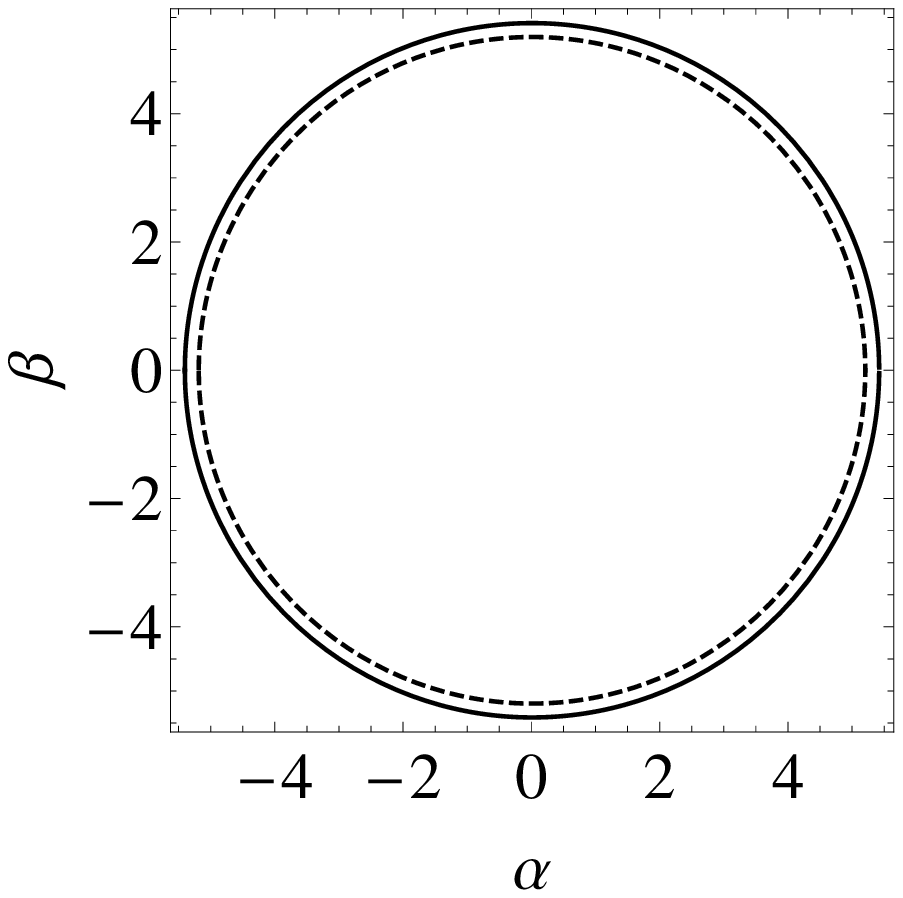} \\
            $a/M_{phys}=0.01$, $\zeta=0.64$;\  &
            $a/M_{phys}=0.01$, $\zeta=0.76$;\  &
            $a/M_{phys}=0.01$, $\zeta=0.88$;\  &
            $a/M_{phys}=0.01$, $\zeta=0.96$ \\
            \includegraphics[width=4.1cm]{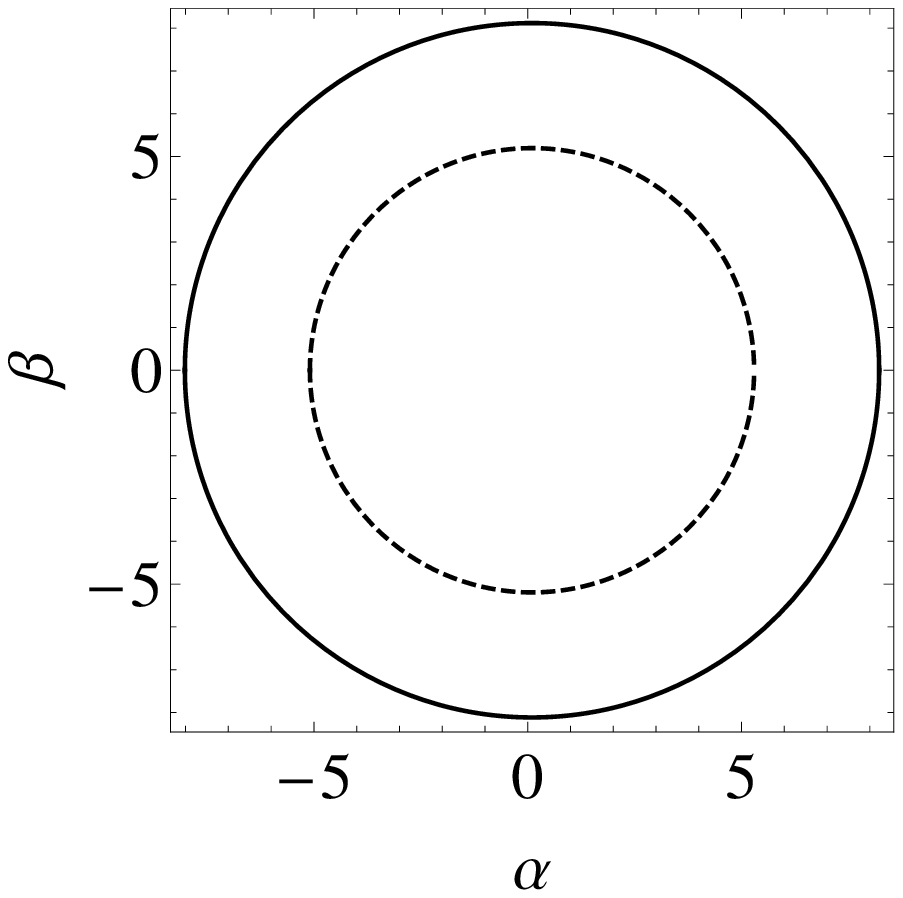} &
            \includegraphics[width=4.1cm]{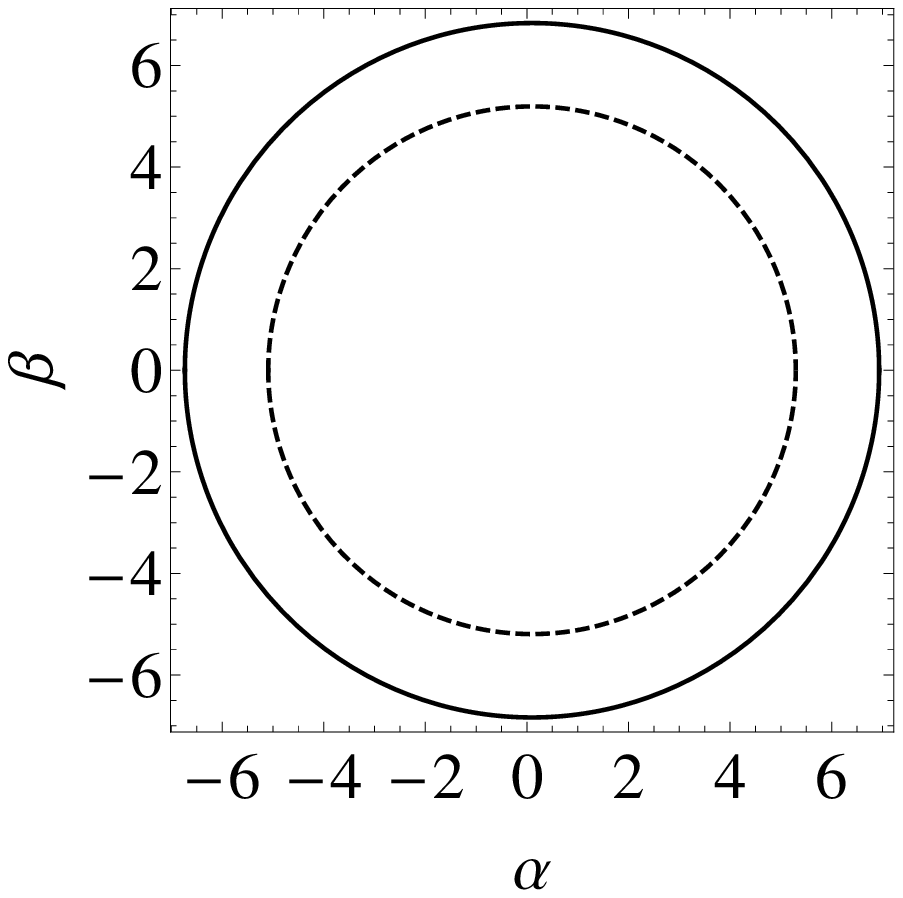} &
            \includegraphics[width=4.1cm]{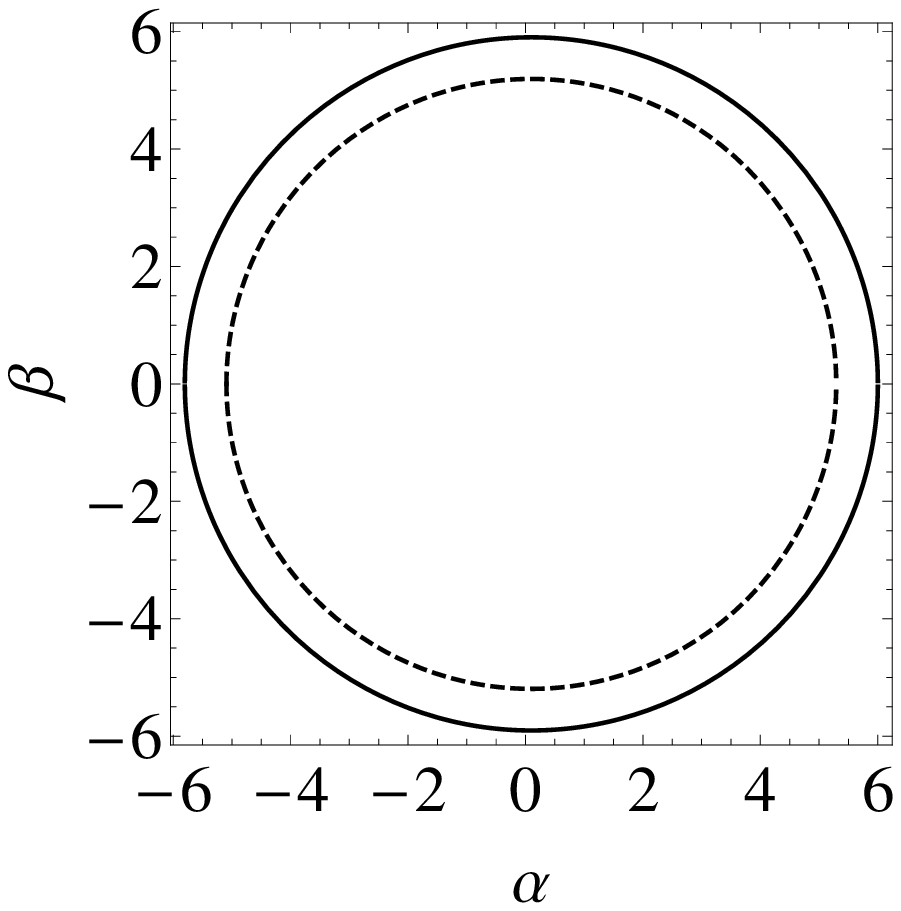} &
            \includegraphics[width=4.1cm]{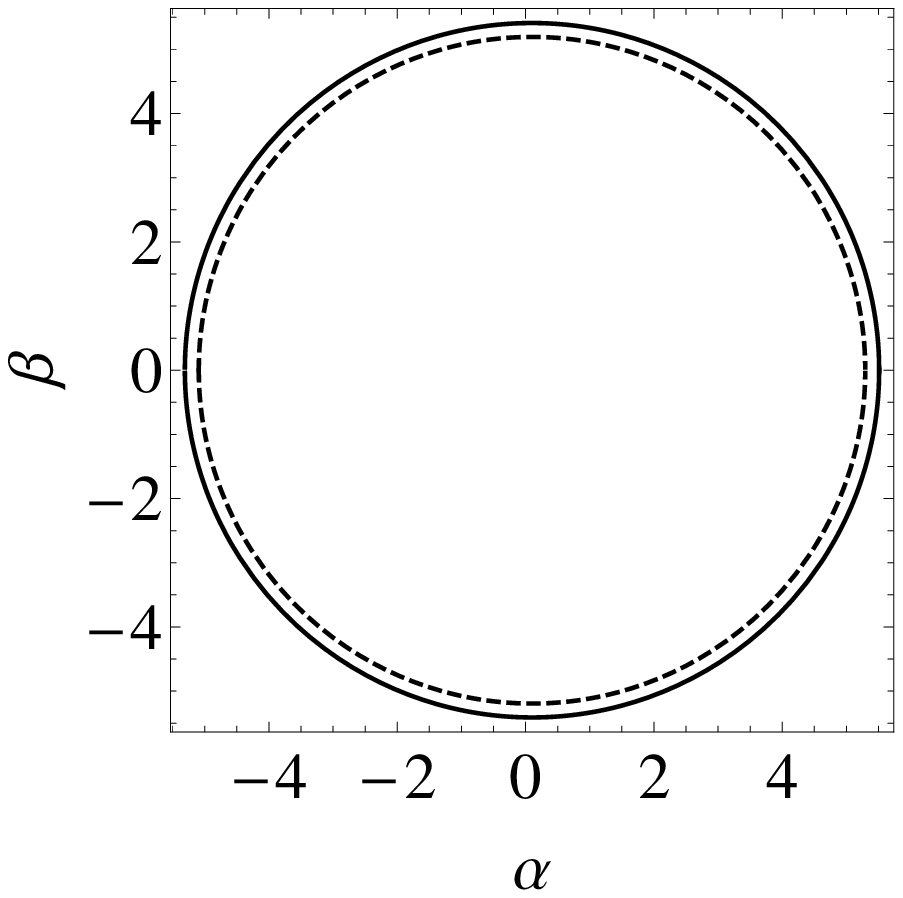} \\
            $a/M_{phys}=0.1$, $\zeta=0.64$;\  &
            $a/M_{phys}=0.1$, $\zeta=0.76$;\  &
            $a/M_{phys}=0.1$, $\zeta=0.88$;\  &
            $a/M_{phys}=0.1$, $\zeta=0.96$ \\
            \includegraphics[width=4.1cm]{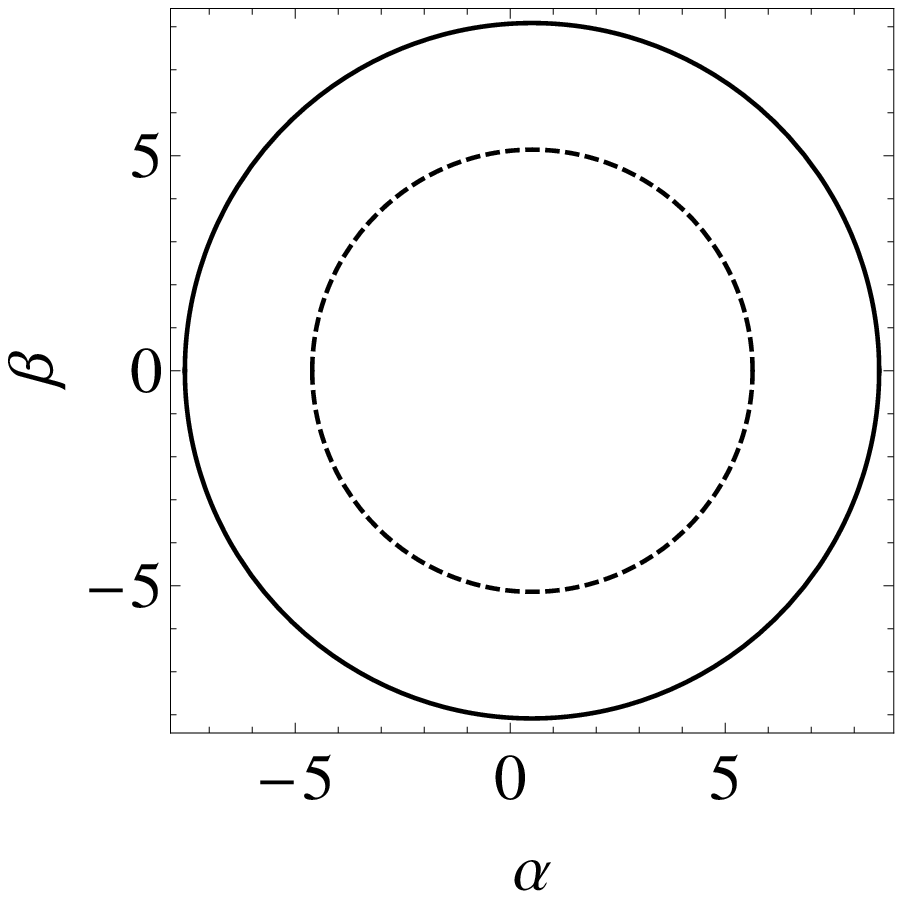} &
            \includegraphics[width=4.1cm]{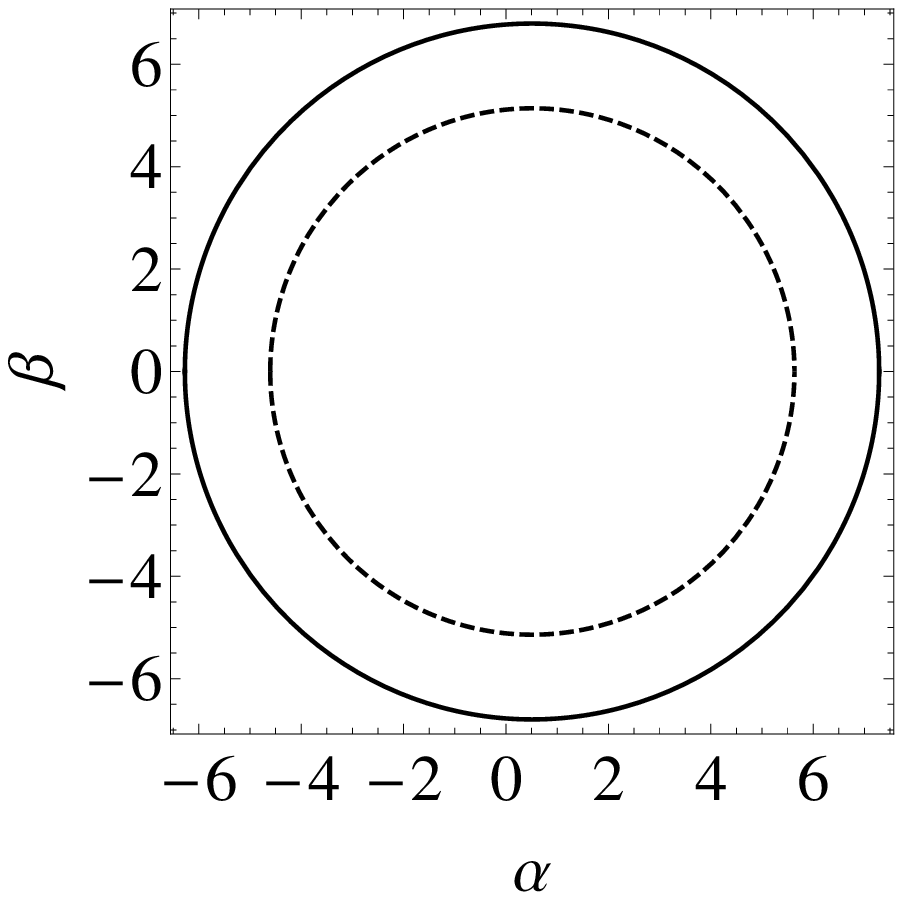} &
            \includegraphics[width=4.1cm]{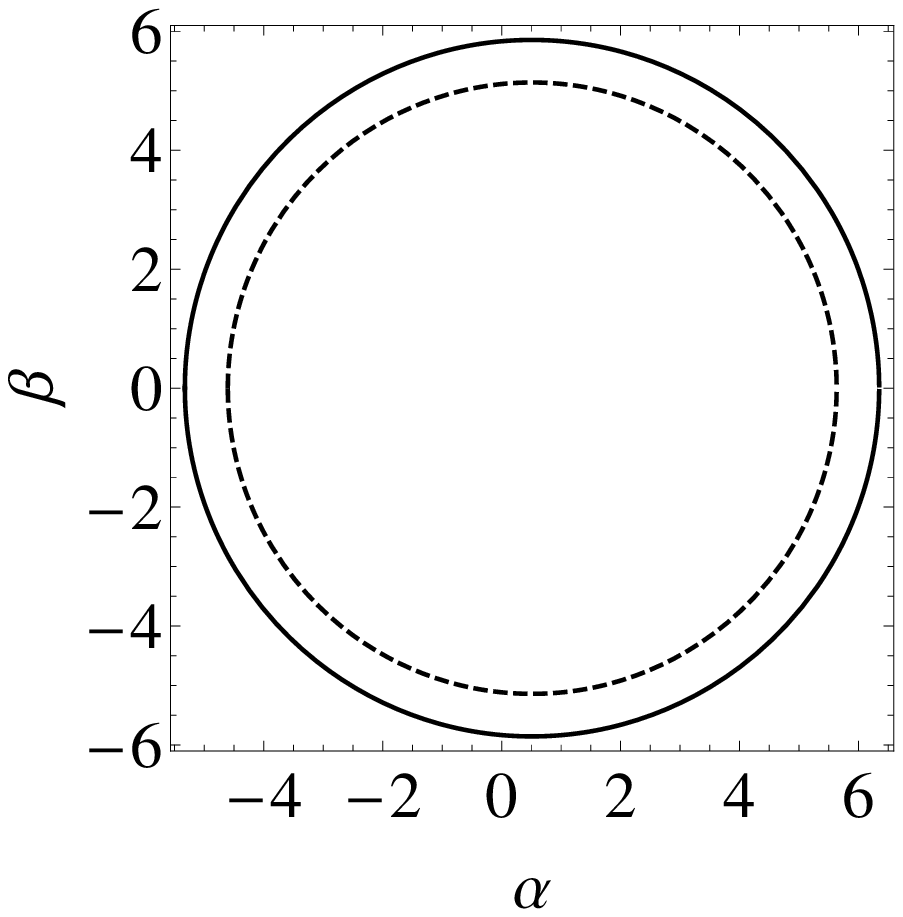} &
            \includegraphics[width=4.1cm]{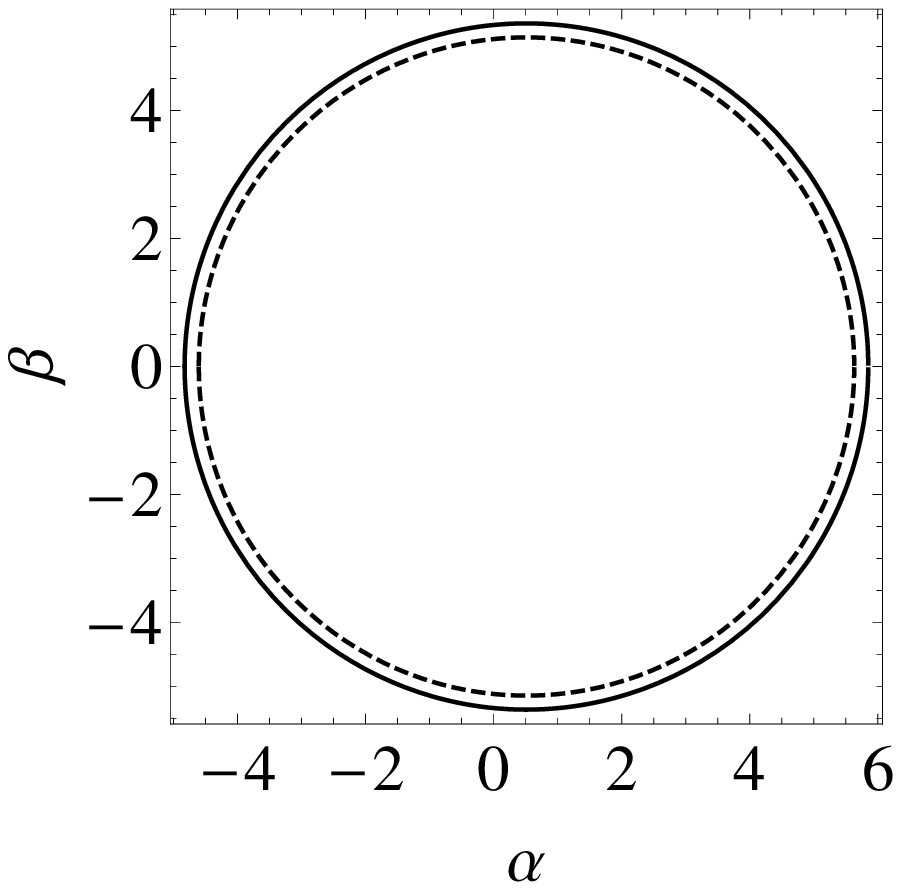} \\
            $a/M_{phys}=0.5$, $\zeta=0.64$;\  &
            $a/M_{phys}=0.5$, $\zeta=0.76$;\  &
            $a/M_{phys}=0.5$, $\zeta=0.88$;\  &
            $a/M_{phys}=0.5$, $\zeta=0.96$ \\
            \includegraphics[width=4.1cm]{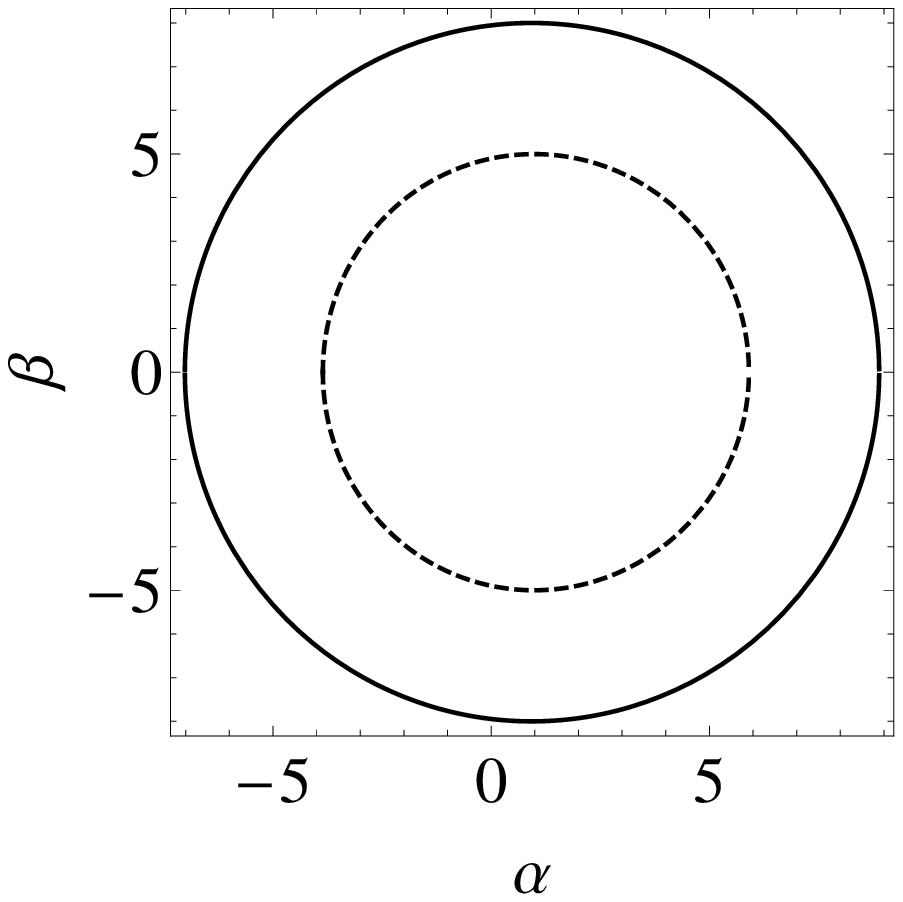} &
            \includegraphics[width=4.1cm]{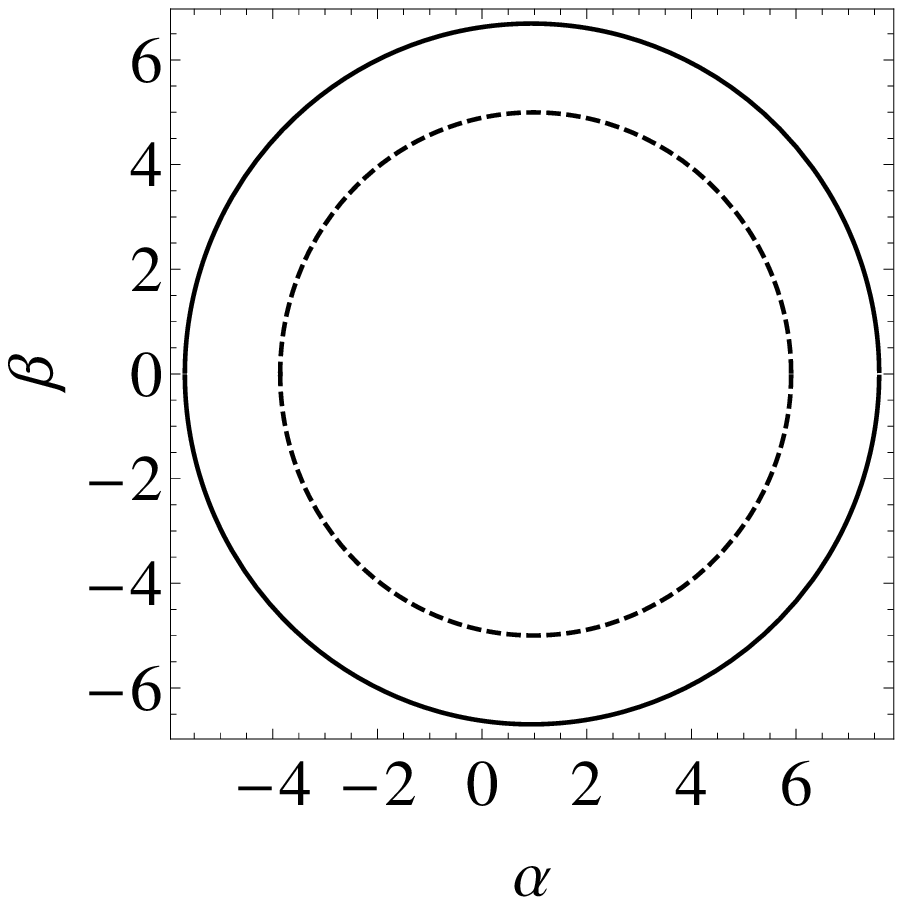} &
            \includegraphics[width=4.1cm]{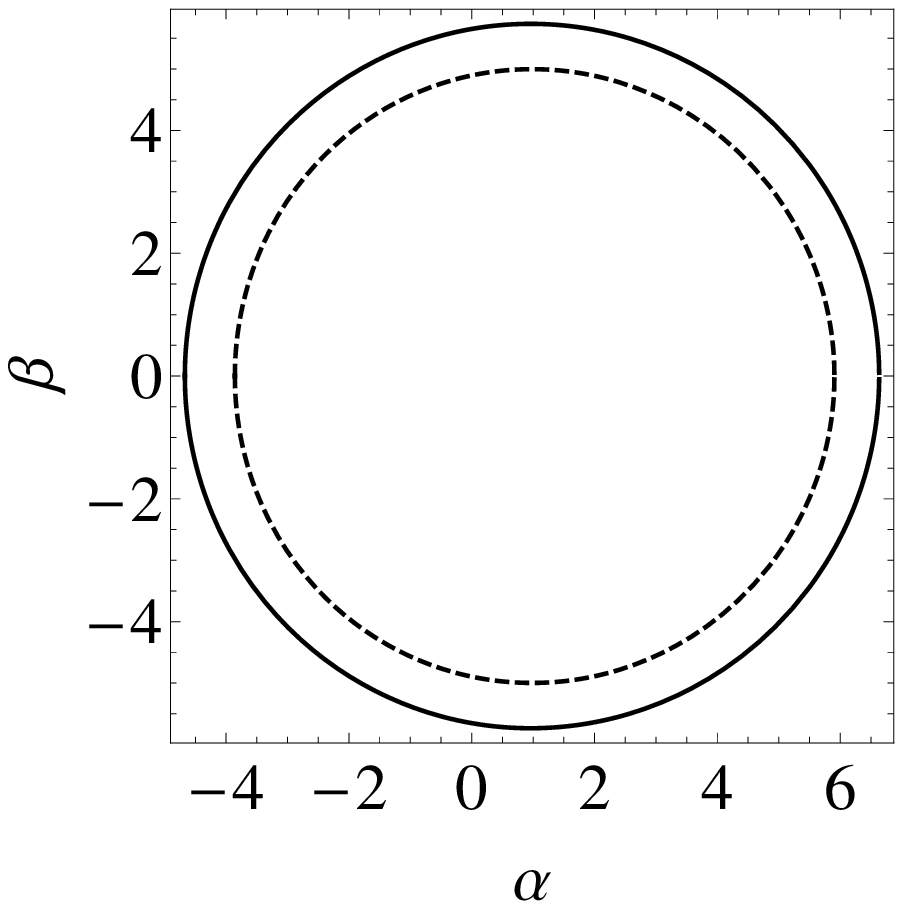} &
            \includegraphics[width=4.1cm]{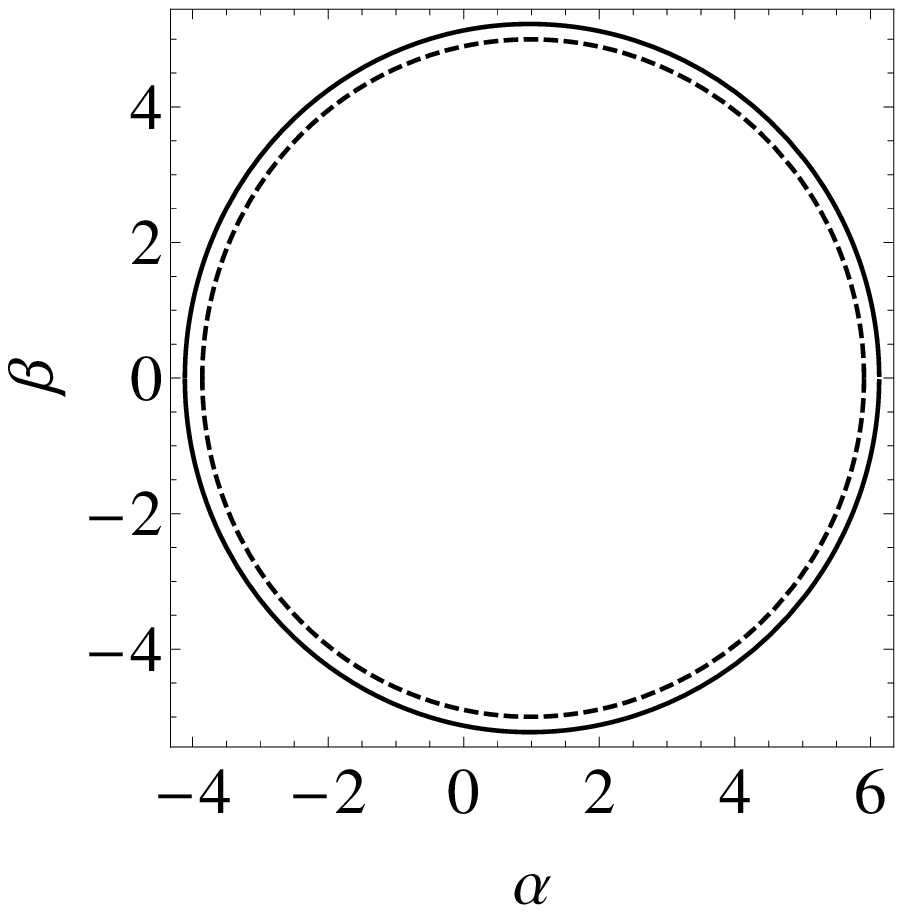} \\
            $a/M_{phys}=0.9$, $\zeta=0.64$;\  &
            $a/M_{phys}=0.9$, $\zeta=0.76$;\  &
            $a/M_{phys}=0.9$, $\zeta=0.88$;\  &
            $a/M_{phys}=0.9$, $\zeta=0.96$ \\
        \end{tabular}}
\caption{\footnotesize{The shadow of the Kerr black hole pierced by
a cosmic string (solid line) and the Kerr black hole (dashed line)
with inclination angle $\theta_{0}=\pi/6\ rad$ for different
rotation and string parameters. The physical mass of both solutions
is set equal to 1. The celestial coordinates $(\alpha,\beta)$ are
measured in the units of physical mass. } }
        \label{WS_a3}
\end{figure}

The shadow of the Kerr black hole pierced by a cosmic string  is
presented in Figs. \ref{WS_a0}--\ref{WS_a3}   for
several inclinations angles, spin and string parameters.

\begin{figure}[h]
    \includegraphics[width=5.0cm]{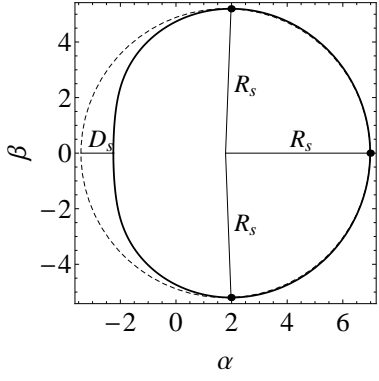}
\caption{\footnotesize{The shadow of Kerr black hole (solid line)
and the circle (dashed line) passing through the three points
located at the top, bottom and rightmost  end of the shadow. The
radius of this circle is $R_{s}$. The difference between the left
end points of the circle and the black hole's shadow is $D_{s}$. The
definition of the distortion parameter is $\delta_{s}\equiv
D_{s}/R_{s}$. } }
        \label{WSCircle_a0}
\end{figure}

There are two observables that characterize  the shadow, namely the
radius $R_{s}$ of the shadow and the so-called dent $D_{s}$
\cite{bibl2}. The radius of the shadow is defined as the circle
passing through the three points located at the top, bottom and
right end of the shadow (see Fig. \ref{WSCircle_a0}). The other
important characteristic, the dent $D_{s}$, is defined as the
difference between the left end points of the circle and the shadow.
It is also useful to define a distortion as $\delta_{s}\equiv
D_{s}/R_{s}$ (see Fig. \ref{WSCircle_a0}).

\begin{figure}[h]
        \setlength{\tabcolsep}{ 0 pt }{\scriptsize\tt
        \begin{tabular}{ cc }
            \includegraphics[width=6.0cm]{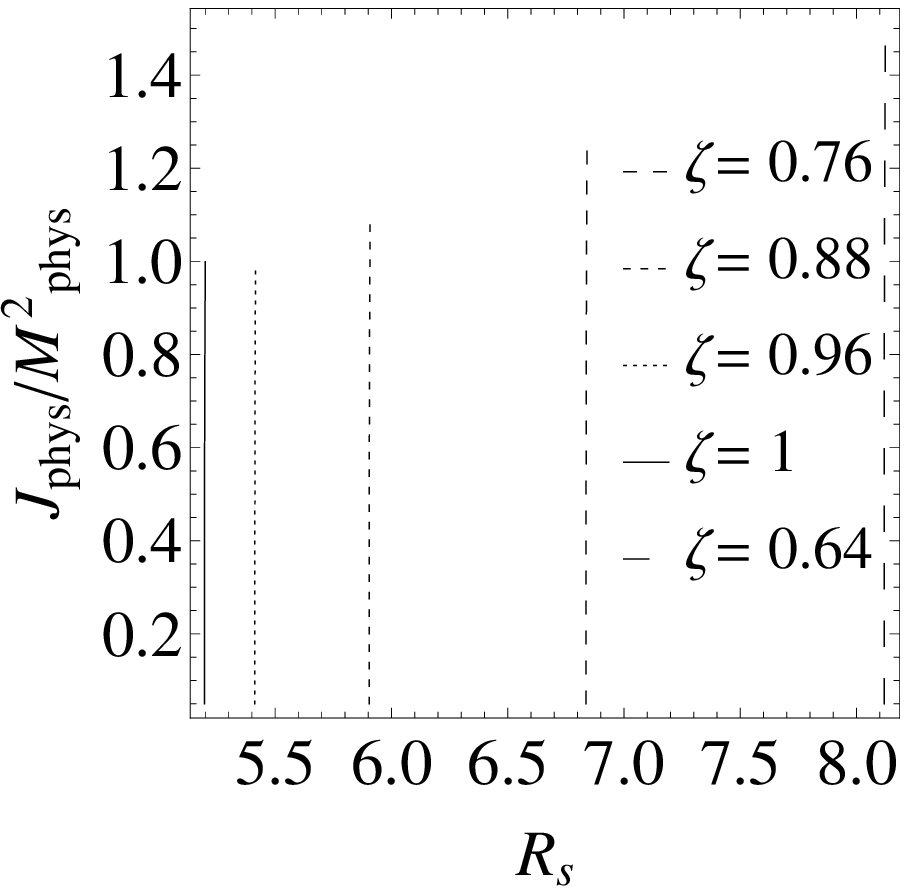} &
            \includegraphics[width=6.0cm]{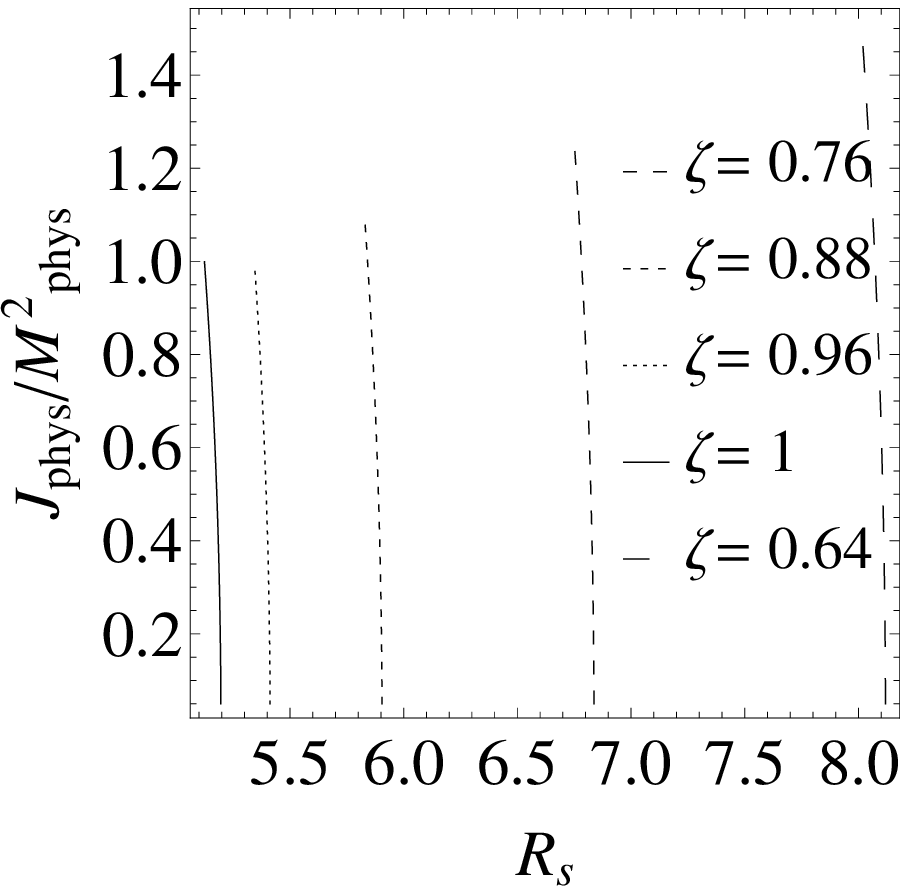} \\
            $\theta_{0}=\pi/2\ rad$;\  &
            $\theta_{0}=\pi/3\ rad$;\  \\
            \includegraphics[width=6.0cm]{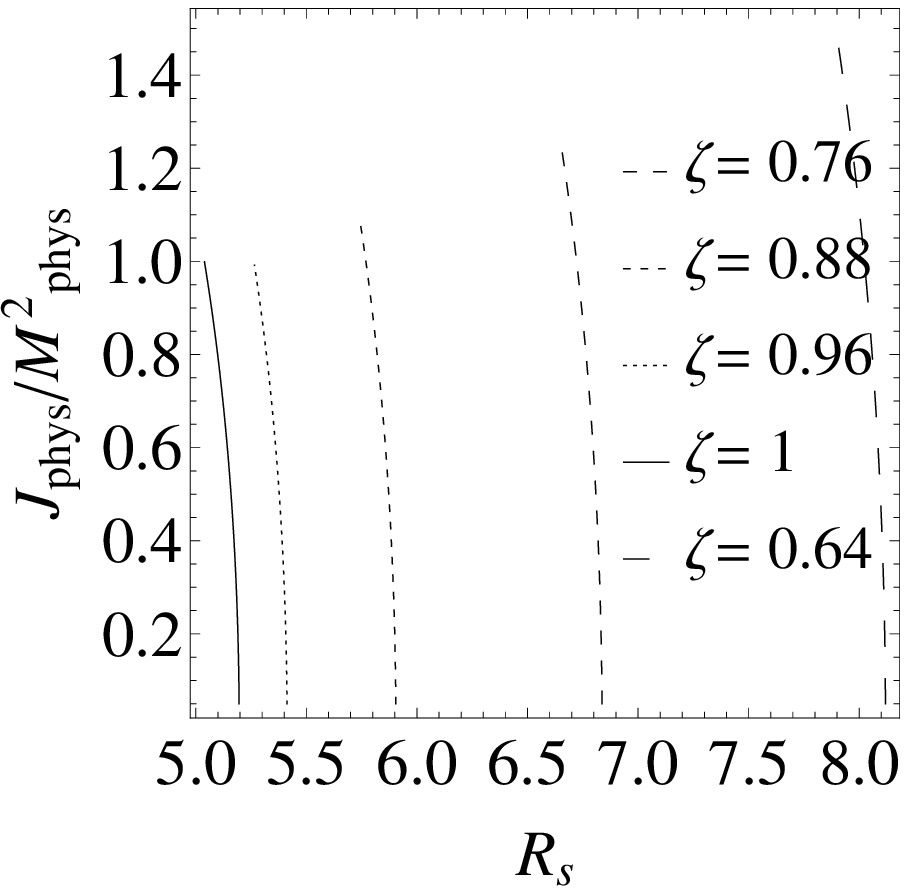} &
            \includegraphics[width=6.0cm]{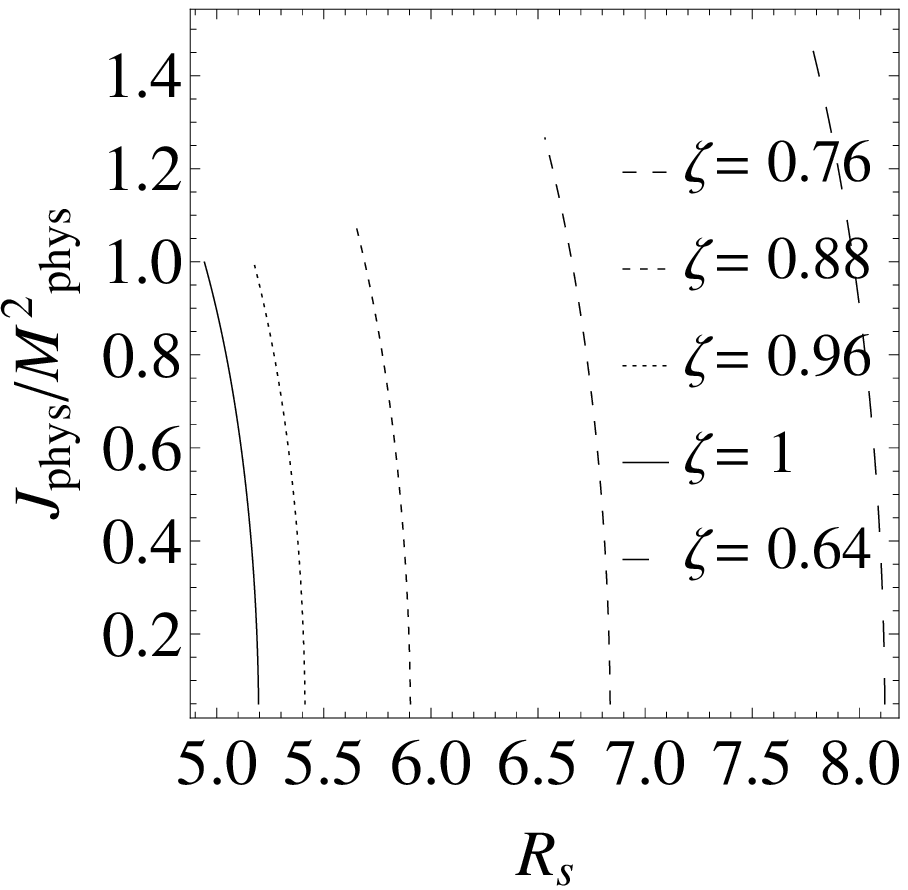} \\
            $\theta_{0}=\pi/4\ rad$;\  &
            $\theta_{0}=\pi/6\ rad$\\
        \end{tabular}}
\caption{\footnotesize{The spin parameter $a_{*}$ of the Kerr black
hole pierced by a cosmic string as a function of the circle radius
$R_{s}$ for a few different inclination angles $\theta_{0}$ and
string parameters $\zeta$. } }
        \label{WSCircle_a2}
\end{figure}

\begin{figure}[h]
        \setlength{\tabcolsep}{ 0 pt }{\scriptsize\tt
        \begin{tabular}{ cc }
            \includegraphics[width=6.0cm]{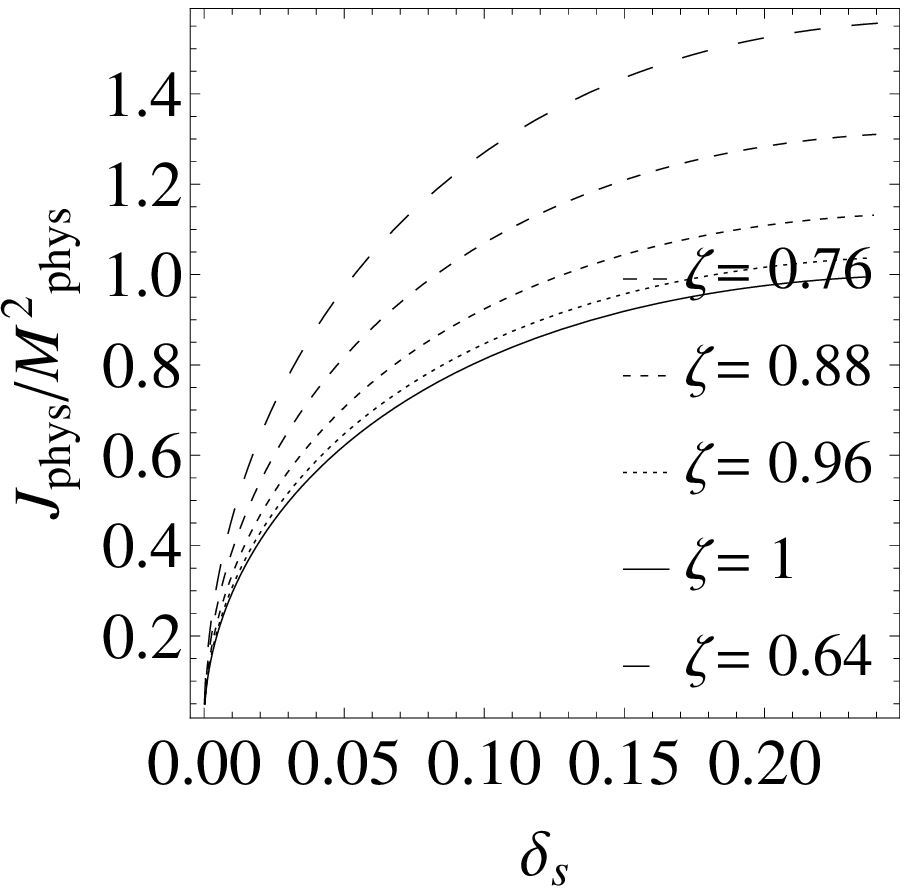} &
            \includegraphics[width=6.0cm]{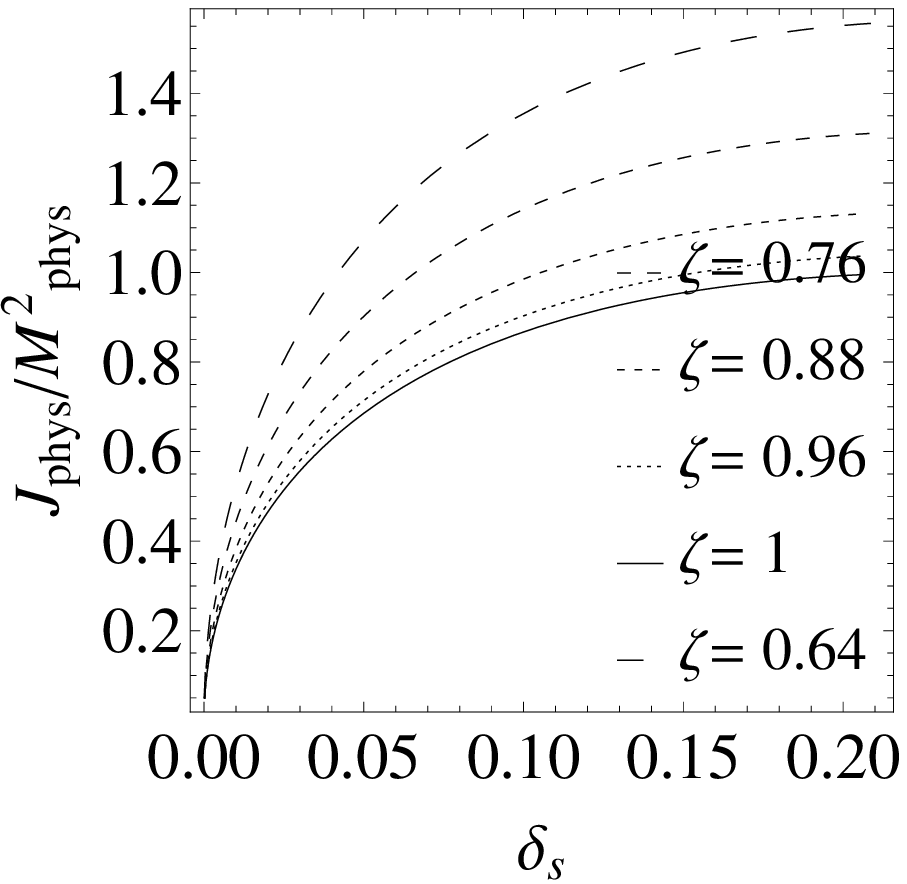} \\
            $\theta_{0}=\pi/2\ rad$;\  &
            $\theta_{0}=\pi/3\ rad$;\  \\
            \includegraphics[width=6.0cm]{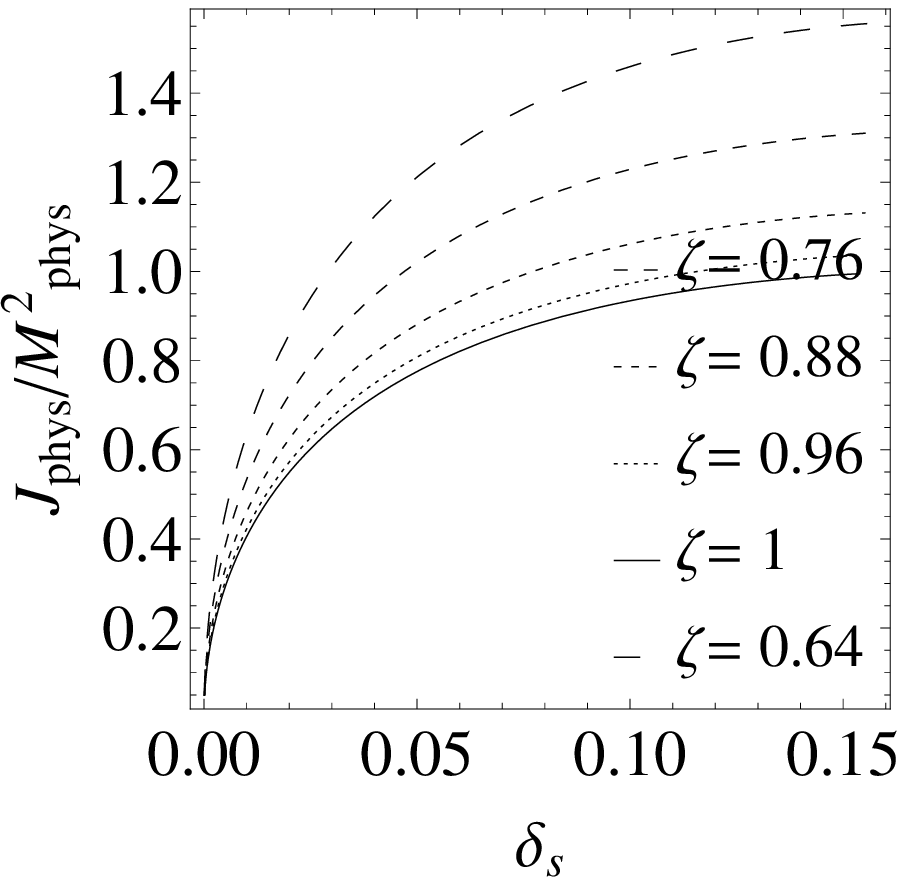} &
            \includegraphics[width=6.0cm]{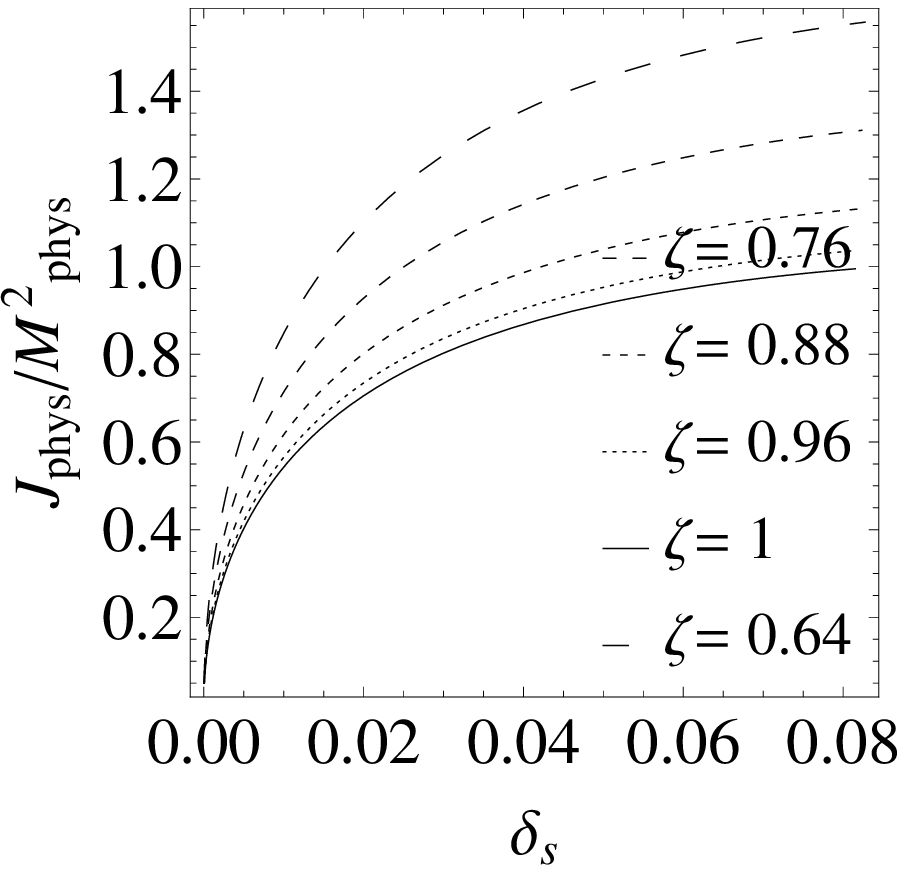} \\
            $\theta_{0}=\pi/4\ rad$;\  &
            $\theta_{0}=\pi/6\ rad$\\
        \end{tabular}}
\caption{\footnotesize{The spin parameter $a_{*}$ of the Kerr black
hole pierced by a cosmic string as a function of the distortion
parameter $\delta_{s}$ for a few different inclination angles
$\theta_{0}$ and string parameters $\zeta$. } }
        \label{WSCircle_a1}
\end{figure}

For small deficit angles (i.e. for $\zeta$ close to $1$) the shadow
of the black hole is very close to that of the Kerr black hole.
However with the increase of the deficit angle the radius of the
shadow boundary increases as it is seen in  Fig. \ref{WSCircle_a2}.
The characteristic deformations of the shadow  are also more
pronounced for large deficit angles as one can see in Fig.
\ref{WSCircle_a1}.

In general, for a fixed mass $M_{phys}$,  the shadow radius $R_{s}$
and the distortion $\delta_{s}$  depend on three parameters - the
spin parameter $a_{*}=J_{phys}/M^2_{phys}$, the string parameter
$\zeta$ and the inclination angle $\theta_{0}$. For fixed
inclination angles, the dependence of $R_{s}$ on the spin parameter
$a_{*}$ for different values of the string parameter $\zeta$ is
shown in Fig. \ref{WSCircle_a2}.  It is seen that the radius $R_{s}$
depends weakly on the spin parameter, while the dependence on the
string parameter is more strongly expressed.

For fixed $M_{phys}$ and $\theta_{0}$ the dependence of the
distortion $\delta_{s}$ on the spin parameter for different values
of the string parameter is shown in Fig.\ref{WSCircle_a1}

All these  results  show that the presence of a cosmic string
piercing the black holes leaves observable imprints in the shadows
of the black holes.  If we can measure the mass, the spin parameter
and the inclination, then our results and more precisely the
dependences shown in Fig. \ref{WSCircle_a2} and Fig.
\ref{WSCircle_a1},  allow for a determination of the string
parameter $\zeta$ which is equivalent to the detection of the
presence of a cosmic string  if $\zeta \ne 0$.

\section*{Acknowledgements}
The support by the Bulgarian National Science Fund under Grant
DMU-03/6 and by the Sofia University Research Fund under Grant
33/2013 is gratefully  acknowledged.

\end{document}